\documentclass[journal,draftclsnofoot,onecolumn,12pt,oneside]{IEEEtran} 

\usepackage{cite}

\ifCLASSINFOpdf
  \usepackage[pdftex]{graphicx}
  \usepackage{epstopdf} 
\else
\fi
\usepackage{amsmath}
\usepackage{algorithm}
\usepackage{algorithmic}

  \usepackage[caption=false,font=footnotesize]{subfig}
\usepackage{url}

\usepackage{amsthm}

\newtheorem{theorem}{Theorem}

\newtheoremstyle{mydefinition}
{}
{}
{}
{0pt}
{\bfseries}
{.}
{ }
{\thmname{#1}\thmnumber{ #2}: \thmnote{#3}}

\theoremstyle{mydefinition}
\newtheorem{definition}{Definition}

\newtheoremstyle{myremark}
{}
{}
{}
{0pt}
{\bfseries}
{.}
{ }
{\thmname{#1}\thmnumber{ #2}: \thmnote{#3}}

\theoremstyle{theoremdd}

\theoremstyle{myremark}
\newcounter{rem}
\newtheorem{remark}[rem]{Remark}

\newcommand{\comment}[1]{{}}

\usepackage{amssymb}
\usepackage{xspace}
\usepackage{bm}
\usepackage{bbm}

\newcommand{\set}[1]{{\mathcal{#1}}\xspace} 
\newcommand{\mat}[1]{{\mathbf{#1}}\xspace} 
\renewcommand{\vec}[1]{{\mathbf{#1}}\xspace} 

\newcommand{\parens}[1]{{\left(#1\right)}\xspace}
\newcommand{\brackets}[1]{{\left[#1\right]}\xspace}
\newcommand{\braces}[1]{{\left\{#1\right\}}\xspace}
\newcommand{\bars}[1]{{\left\vert#1\right\vert}\xspace}
\newcommand{\doublebars}[1]{{\left\Vert#1\right\Vert}\xspace}

\newcommand{\complex}{\ensuremath{\mathbb{C}}\xspace}

\renewcommand{\j}{\ensuremath{\mathrm{j}}}

\newcommand{\setcomplex}{\ensuremath{\complex}}

\newcommand{\setvector}[2]{\ensuremath{#1^{#2 \times 1}}\xspace}

\newcommand{\setvectorcomplex}[1]{\setvector{\setcomplex}{#1}}

\newcommand{\setmatrix}[3]{\ensuremath{#1^{#2 \times #3}}\xspace}

\newcommand{\setmatrixcomplex}[2]{\setmatrix{\setcomplex}{#1}{#2}}

\newcommand{\zeromat}{\ensuremath{\mat{0}}\xspace}

\newcommand{\ctrans}{\ensuremath{^{{*}}}\xspace}

\newcommand{\trace}[1]{\ensuremath{\mathrm{tr}\parens{#1}}\xspace}

\newcommand{\entry}[2]{\ensuremath{\brackets{#1}_{#2}}\xspace}

\newcommand{\logtwo}[1]{\ensuremath{\mathrm{log}_{2}\parens{#1}}}

\newcommand{\diag}[1]{\ensuremath{\mathrm{diag}\parens{#1}}}

\newcommand{\angleop}[1]{\ensuremath{\mathrm{angle}\parens{#1}}}

\newcommand{\svmax}[1]{\ensuremath{\sigma_{\mathrm{max}}\parens{#1}}\xspace}

\newcommand{\svmaxsq}[1]{\ensuremath{\sigma_{\mathrm{max}}^2\parens{#1}}\xspace}

\newcommand{\pnorm}[2]{\ensuremath{\doublebars{#2}_{#1}}\xspace}

\newcommand{\normtwo}[1]{\pnorm{2}{#1}}

\newcommand{\normfro}[1]{\pnorm{\mathrm{F}}{#1}}

\newcommand{\distcgauss}[2]{\ensuremath{\mathcal{N}_{\complex}\parens{#1,#2}}\xspace} 

\newcommand{\ev}[1]{\ensuremath{\mathbb{E}\brackets{#1}}\xspace}
\DeclareMathOperator*{\expect}{\mathbb{E}\xspace}

\DeclareMathOperator*{\argmin}{argmin}

\newcommand{\maxop}[1]{\ensuremath{\mathrm{max}\parens{#1}}\xspace}

\newcommand{\st}{\ensuremath{\mathrm{s.t.~}}\xspace}
\newcommand{\opt}{\ensuremath{^{\star}}\xspace}
\newcommand{\project}[2]{\ensuremath{\Pi_{#2}\parens{#1}}\xspace}

\newcommand{\powernoise}{\ensuremath{P_{\mathrm{noise}}}\xspace}

\newcommand{\powertx}{\ensuremath{P_{\mathrm{tx}}}\xspace}

\newcommand{\powertxue}{\ensuremath{\powertx^{\mathrm{UE}}}\xspace}
\newcommand{\powertxbs}{\ensuremath{\powertx^{\mathrm{BS}}}\xspace}

\newcommand{\snr}{\ensuremath{\mathsf{SNR}}\xspace}

\newcommand{\sinr}{\ensuremath{\mathsf{SINR}}\xspace}

\newcommand{\inr}{\ensuremath{\mathsf{INR}}\xspace}

\newcommand{\atx}[1]{\ensuremath{\vec{a}_{\mathrm{tx}}\parens{#1}}\xspace}
\newcommand{\arx}[1]{\ensuremath{\vec{a}_{\mathrm{rx}}\parens{#1}}\xspace}

\newcommand{\Atx}{\ensuremath{\mA_{\mathrm{tx}}}\xspace}
\newcommand{\Arx}{\ensuremath{\mA_{\mathrm{rx}}}\xspace}

\newcommand{\Nt}{\ensuremath{N_\mathrm{t}}\xspace} 
\newcommand{\Nr}{\ensuremath{N_\mathrm{r}}\xspace} 
\newcommand{\Na}{\ensuremath{N_{\mathrm{a}}}\xspace} 

\newcommand{\precbmat}{\ensuremath{\mat{F}}\xspace}
\newcommand{\comcbmat}{\ensuremath{\mat{W}}\xspace}

\newcommand{\pre}{\ensuremath{\mat{F}}\xspace}

\newcommand{\prerf}{\ensuremath{\pre_{\mathrm{RF}}}\xspace}

\newcommand{\precb}{\ensuremath{\mathcal{F}}\xspace}

\newcommand{\prerfcb}{\ensuremath{\mathcal{F}_{\mathrm{RF}}}\xspace}

\newcommand{\com}{\ensuremath{\mat{W}}\xspace}

\newcommand{\comrf}{\ensuremath{\com_{\mathrm{RF}}}\xspace}

\newcommand{\comcb}{\ensuremath{\mathcal{W}}\xspace}

\newcommand{\comrfcb}{\ensuremath{\mathcal{W}_{\mathrm{RF}}}\xspace}

\newcommand{\sigmatx}{\ensuremath{\sigma_{\mathrm{tx}}}\xspace}
\newcommand{\sigmarx}{\ensuremath{\sigma_{\mathrm{rx}}}\xspace}

\newcommand{\bitsamp}{\ensuremath{{b_{\mathrm{amp}}}}\xspace}
\newcommand{\bitsphase}{\ensuremath{{b_{\mathrm{phs}}}}\xspace}

\newcommand{\prerfcbbar}{\ensuremath{\bar{\mathcal{F}}_{\mathrm{RF}}}\xspace}
\newcommand{\comrfcbbar}{\ensuremath{\bar{\mathcal{W}}_{\mathrm{RF}}}\xspace}

\newcommand{\idx}[1]{\ensuremath{^{\parens{#1}}}\xspace}

\newcommand{\txdirsetcb}{\ensuremath{\set{A}_{\mathrm{tx}}}\xspace}
\newcommand{\rxdirsetcb}{\ensuremath{\set{A}_{\mathrm{rx}}}\xspace}

\newcommand{\thetatx}{\ensuremath{\theta_{\mathrm{tx}}}\xspace}
\newcommand{\phitx}{\ensuremath{\phi_{\mathrm{tx}}}\xspace}

\newcommand{\thetarx}{\ensuremath{\theta_{\mathrm{rx}}}\xspace}
\newcommand{\phirx}{\ensuremath{\phi_{\mathrm{rx}}}\xspace}

\newcommand{\Mtx}{\ensuremath{M_{\mathrm{tx}}}\xspace}
\newcommand{\Mrx}{\ensuremath{M_{\mathrm{rx}}}\xspace}

\newcommand{\labelue}{\mathrm{UE}}
\newcommand{\labeltx}{\mathrm{tx}}
\newcommand{\labelrx}{\mathrm{rx}}
\newcommand{\labelsum}{\mathrm{sum}}

\newcommand{\labelcb}{\mathrm{cb}}

\newcommand{\snrtx}{\ensuremath{\snr_{\labeltx}}\xspace}
\newcommand{\snrrx}{\ensuremath{\snr_{\labelrx}}\xspace}

\newcommand{\sinrtx}{\ensuremath{\sinr_{\labeltx}}\xspace}
\newcommand{\sinrrx}{\ensuremath{\sinr_{\labelrx}}\xspace}

\newcommand{\inrtx}{\ensuremath{\inr_{\labeltx}}\xspace}
\newcommand{\inrrx}{\ensuremath{\inr_{\labelrx}}\xspace}
\newcommand{\inrrxthresh}{\ensuremath{\inr_{\labelrx}^{\mathrm{tgt}}}\xspace}

\newcommand{\snrtxbar}{\ensuremath{\overline{\snr}_{\labeltx}}\xspace}
\newcommand{\snrrxbar}{\ensuremath{\overline{\snr}_{\labelrx}}\xspace}

\newcommand{\captx}{\ensuremath{C_{\labeltx}}\xspace}
\newcommand{\caprx}{\ensuremath{C_{\labelrx}}\xspace}

\newcommand{\captxcb}{C_{\labeltx}^{\labelcb}\xspace}
\newcommand{\caprxcb}{C_{\labelrx}^{\labelcb}\xspace}

\newcommand{\setx}{\ensuremath{R_{\labeltx}}\xspace}
\newcommand{\serx}{\ensuremath{R_{\labelrx}}\xspace}

\newcommand{\hcl}{h_{\mathrm{CL}}\xspace}

\newcommand{\vhtx}{\vh_{\labeltx}\xspace}
\newcommand{\vhrx}{\vh_{\labelrx}\xspace}

\newcommand{\Gtx}{\ensuremath{G_{\labeltx}}\xspace}
\newcommand{\Grx}{\ensuremath{G_{\labelrx}}\xspace}

\newcommand{\Gcl}{\ensuremath{G_{\mathrm{CL}}}\xspace}

\newcommand{\powernoiseue}{\ensuremath{\powernoise^{\labelue}}\xspace}
\newcommand{\powernoisebs}{\ensuremath{\powernoise^{\mathrm{BS}}}\xspace}

\newcommand{\mHbar}{\bar{\mH}\xspace}
\newcommand{\eps}{\epsilon\xspace}

\newcommand{\inrrxbar}{\ensuremath{\overline{\inr}_{\labelrx}}\xspace}
\newcommand{\dir}{\ensuremath{\theta}\xspace}
\newcommand{\dirtx}{\ensuremath{\dir_{\mathrm{tx}}}\xspace}
\newcommand{\dirrx}{\ensuremath{\dir_{\mathrm{rx}}}\xspace}
\newcommand{\dirset}{\ensuremath{\mathcal{A}}\xspace}
\newcommand{\dirtxset}{\ensuremath{\mathcal{A}_{\mathrm{tx}}}\xspace}
\newcommand{\dirrxset}{\ensuremath{\mathcal{A}_{\mathrm{rx}}}\xspace}
\newcommand{\precbbar}{\ensuremath{\bar{\mathcal{F}}}\xspace}
\newcommand{\comcbbar}{\ensuremath{\bar{\mathcal{W}}}\xspace}
\newcommand{\inrrxavg}{\ensuremath{\inr_{\labelrx}^{\mathrm{avg}}}\xspace}
\newcommand{\sesumnorm}{\ensuremath{\gamma_{\labelsum}}\xspace}

\newcommand{\labelcbf}{\mathrm{CBF}}
\newcommand{\snrtxcbf}{\ensuremath{\snr_{\labeltx}^{\labelcbf}}\xspace}
\newcommand{\snrrxcbf}{\ensuremath{\snr_{\labelrx}^{\labelcbf}}\xspace}
\newcommand{\sigmatxsq}{\ensuremath{\sigmatx^2}\xspace}
\newcommand{\sigmarxsq}{\ensuremath{\sigmarx^2}\xspace}
\newcommand{\sigmatxrxsq}{\ensuremath{\sigmatxsq = \sigmarxsq}\xspace}
\newcommand{\bitsphaseamp}{\ensuremath{\bitsphase = \bitsamp}\xspace}
\newcommand{\snrtxrxbar}{\ensuremath{\snrtxbar = \snrrxbar}\xspace}

\def\vf{{\vec{f}}}

\def\vh{{\vec{h}}}

\def\vv{{\vec{v}}}
\def\vw{{\vec{w}}}
\def\vx{{\vec{x}}}

\def\vone{{\vec{1}}}

\def\mA{{\mat{A}}}

\def\mF{{\mat{F}}}

\def\mH{{\mat{H}}}
\def\mI{{\mat{I}}}

\def\mU{{\mat{U}}}

\def\mW{{\mat{W}}}
\def\mX{{\mat{X}}}

\def\mDelta{{\mat{\Delta}}}

\def\mLambda{{\mat{\Lambda}}}

\usepackage{xspace}
\usepackage[acronym,nogroupskip,nonumberlist,nopostdot]{glossaries}
\makeglossaries

\newcommand{\steer}{\textsc{Steer}\xspace}
\newcommand{\lonestar}{\textsc{LoneSTAR}\xspace}
\newcommand{\cbf}{CBF\xspace}
\newcommand{\taylor}{Taylor\xspace}

\usepackage{xcolor}
\newcommand{\edit}[1]{\textcolor{black}{#1}}

\newacronym{snr}{SNR}{signal-to-noise ratio}
\newacronym{sinr}{SINR}{signal-to-interference-plus-noise ratio}
\newacronym{inr}{INR}{interference-to-noise ratio}
\newacronym{sir}{SIR}{signal-to-interference ratio}
\newacronym{sqr}{SQR}{signal-to-quantization-noise ratio}
\newacronym{sqnr}{SQNR}{signal-to-quantization-plus-noise ratio}
\newacronym{ian}{IAN}{interference as noise}
\newacronym{ber}{BER}{bit error rate}
\newacronym{pn}{PN}{pseudorandom noise}
\newacronym{bfsk}{BFSK}{binary frequency shift keying}
\newacronym{fh}{FH}{frequency-hopped}
\newacronym{fh-bfsk}{FH-BFSK}{frequency-hopped binary frequency shift keying}
\newacronym{crc}{CRC}{cyclic redundancy check}
\newacronym{isi}{ISI}{intersymbol interference}
\newacronym{dsss}{DSSS}{direct-sequence spread spectrum}
\newacronym{ofdm}{OFDM}{orthogonal frequency-division multiplexing}
\newacronym{ofdma}{OFDMA}{orthogonal frequency-division multiple access}
\newacronym{sdr}{SDR}{software-defined radio}
\newacronym{tx}{TX}{transmitter}
\newacronym{rx}{RX}{receiver}
\newacronym{fdd}{FDD}{frequency-division duplexing}
\newacronym{tdd}{TDD}{time-division duplexing}
\newacronym{fdma}{FDMA}{frequency-division multiple access}
\newacronym{tdma}{TDMA}{time-division multiple access}
\newacronym{sdma}{SDMA}{space-division multiple access}
\newacronym[plural=MPCs]{mpc}{MPC}{multipath component}
\newacronym{mui}{MUI}{multi-user interference}
\newacronym{lsb}{LSB}{least significant bit}

\newacronym{qam}{QAM}{quadrature amplitude modulation}
\newacronym{mqam}{MQAM}{M-ary quadrature amplitude modulation}

\newacronym{ls}{LS}{least-squares}
\newacronym{lms}{LMS}{least mean squares}
\newacronym{rls}{RLS}{recursive least-squares}
\newacronym{rzf}{RZF}{regularized zero-forcing}
\newacronym{mmse}{MMSE}{minimum mean square error}
\newacronym{lmmse}{LMMSE}{linear minimum mean square error}
\newacronym{mse}{MSE}{mean square error}
\newacronym{fft}{FFT}{fast Fourier transform}
\newacronym{dft}{DFT}{discrete Fourier transform}
\newacronym{dtft}{DTFT}{discrete-time Fourier transform}
\newacronym{ctft}{CTFT}{continuous-time Fourier transform}
\newacronym{ml}{ML}{machine learning}
\newacronym[plural=NNs]{nn}{NN}{neural network}
\newacronym[plural=RNNs]{rnn}{RNN}{recurrent neural network}
\newacronym[plural=ADCs]{adc}{ADC}{analog-to-digital converter}
\newacronym[plural=DACs]{dac}{DAC}{digital-to-analog converter}
\newacronym[plural=FPGAs]{fpga}{FPGA}{field-programmable gate array}
\newacronym{evm}{EVM}{error vector magnitude}
\newacronym{enob}{ENOB}{effective number of bits}
\newacronym{zf}{ZF}{zero-forcing}
\newacronym{rv}{r.v.}{random variable}
\newacronym{omp}{OMP}{orthogonal matching pursuit}
\newacronym{svd}{SVD}{singular value decomposition}

\newacronym{sdp}{SDP}{semidefinite programming}
\newacronym{psd}{PSD}{positive semidefinite}
\newacronym{nsd}{NSD}{negative semidefinite}

\newacronym{ks}{K-S}{Kolmogorov-Smirnov}

\newacronym{mad}{MAD}{median absolute deviation around the median}

\newacronym{agc}{AGC}{automatic gain control}
\newacronym{rf}{RF}{radio frequency}
\newacronym{if}{IF}{intermediate frequency}
\newacronym{los}{LOS}{line-of-sight}
\newacronym{nlos}{NLOS}{non-line-of-sight}
\newacronym{ple}{PLE}{path loss exponent}
\newacronym[plural=dB,firstplural=decibels (dB)]{db}{dB}{decibel}
\newacronym[plural=dBm,firstplural=decibel milliwatts (dBm)]{dbm}{dBm}{decibel milliwatts}
\newacronym{pa}{PA}{power amplifier}
\newacronym{lna}{LNA}{low-noise amplifier}
\newacronym{vga}{VGA}{variable-gain amplifier}
\newacronym{cw}{CW}{continuous wave}
\newacronym{papr}{PAPR}{peak-to-average power ratio}
\newacronym{usrp}{USRP}{Universal Software Radio Peripheral}
\newacronym{irr}{IRR}{image rejection ratio}
\newacronym{lo}{LO}{local oscillator}
\newacronym{vm}{VM}{vector modulator}
\newacronym{mmwave}{mmWave}{millimeter wave}
\newacronym{eirp}{EIRP}{effective isotropic radiated power}
\newacronym{rsrp}{RSRP}{reference signal received power}

\newacronym{csma}{CSMA}{carrier-sense multiple access}
\newacronym{csmaca}{CSMA/CA}{carrier-sense multiple access with collision avoidance}
\newacronym{csmacd}{CSMA/CD}{carrier-sense multiple access with collision detection}
\newacronym{mac}{MAC}{medium access control}
\newacronym{phy}{PHY}{physical layer}
\newacronym{4g}{4G}{fourth-generation}
\newacronym{lte}{LTE}{Long-Term Evolution}
\newacronym{4glte}{4G LTE}{\gls{4g} \gls{lte}}
\newacronym{5g}{5G}{fifth-generation}
\newacronym{nr}{NR}{New Radio}
\newacronym{5gnr}{5G NR}{5G New Radio}
\newacronym{ieee}{IEEE}{Institute of Electrical and Electronics Engineers}
\newacronym{wifi}{Wi-Fi}{IEEE 802.11}
\newacronym{lan}{LAN}{local area network}
\newacronym{wlan}{WLAN}{wireless local area network}
\newacronym[plural=BSs]{bs}{BS}{base station}
\newacronym[plural=SBSs]{sbs}{SBS}{small-cell base station}
\newacronym[plural=FD-SBSs]{fdsbs}{FD-SBS}{\gls{fd}-enabled \gls{sbs}}
\newacronym[plural=MBSs]{mbs}{MBS}{macrocell base station}
\newacronym[plural=UEs]{ue}{UE}{user equipment}
\newacronym{ul}{UL}{uplink}
\newacronym{dl}{DL}{downlink}
\newacronym{qos}{QoS}{quality of service}
\newacronym{fcc}{FCC}{Federal Communications Commission}
\newacronym{iab}{IAB}{integrated access and backhaul}
\newacronym{fab}{FAB}{fixed access and backhaul}
\newacronym{hetnet}{HetNet}{heterogeneous network}

\newacronym{siso}{SISO}{single-input single-output}
\newacronym{mimo}{MIMO}{multiple-input multiple-output}
\newacronym{sumimo}{SU-MIMO}{single-user \gls{mimo}}
\newacronym{mumimo}{MU-MIMO}{multi-user \gls{mimo}}
\newacronym{bf}{BF}{beamforming}
\newacronym{ca}{CA}{constant amplitude}
\newacronym{ula}{ULA}{uniform linear array}
\newacronym{upa}{UPA}{uniform planar array}
\newacronym[\glslongpluralkey={angles of arrival}]{aoa}{AoA}{angle of arrival}
\newacronym[\glslongpluralkey={angles of departure}]{aod}{AoD}{angle of departure}
\newacronym{dof}{DoF}{degrees of freedom}
\newacronym{csi}{CSI}{channel state information}
\newacronym{csit}{CSIT}{\gls{csi} at the transmitter}
\newacronym{csir}{CSIR}{\gls{csi} at the receiver}
\newacronym{cs}{CS}{compressed sensing}

\newacronym{fd}{FD}{in-band full-duplex}
\newacronym{hd}{HD}{half-duplex}
\newacronym{si}{SI}{self-interference}
\newacronym{sic}{SIC}{self-interference cancellation}
\newacronym{soi}{SoI}{signal of interest}
\newacronym{asic}{A-SIC}{analog \acrlong{sic}}
\newacronym{dsic}{D-SIC}{digital \gls{sic}}
\newacronym{star}{STAR}{simultaneous transmit and receive}
\newacronym{warp}{WARP}{Wireless Open-Access Research Platform}
\newacronym{bfc}{BFC}{beamforming cancellation}
\newacronym{ipi}{IPI}{inter-panel-interference}
\newacronym{ipic}{IPIC}{inter-panel-interference cancellation}

\newacronym{qcqp}{QCQP}{quadratically-constrained quadratic programming}
\newacronym{pdf}{PDF}{probability density function}
\newacronym{cdf}{CDF}{cumulative density function}
\newacronym{iid}{i.i.d.}{independently and identically distributed}

\newacronym{elf}{ELF}{extremely low frequency}
\newacronym{slf}{SLF}{super low frequency}
\newacronym{ulf}{ULF}{ultra low frequency}
\newacronym{vlf}{VLF}{very low frequency}
\newacronym{lf}{LF}{low frequency}
\newacronym{mf}{MF}{medium frequency}
\newacronym{hf}{HF}{high frequency}
\newacronym{vhf}{VHF}{very high frequency}
\newacronym{uhf}{UHF}{ultra high frequency}
\newacronym{shf}{SHF}{super high frequency}
\newacronym{ehf}{EHF}{extremely high frequency}
\newacronym{thf}{THF}{tremendously high frequency}

\newacronym{wncg}{WNCG}{Wireless Networking and Communications Group}
\newacronym{linc}{LINC}{Laboratory of Informatics, Networks, and Communications}
\newacronym{ut}{UT Austin}{The University of Texas at Austin}
\newacronym{uiuc}{UIUC}{University of Illinois at Urbana-Champaign}
\newacronym{usc}{USC}{University of Southern California}
\newacronym{mit}{MIT}{Massachusetts Institute of Technology}
\newacronym{berkeley}{UC Berkeley}{University of California, Berkeley}
\newacronym{osu}{OSU}{Ohio State University}

\newcommand{\dacs}{\glspl{dac}\xspace}
\newcommand{\iid}{\gls{iid}\xspace}

\newcommand{\mmwave}{\gls{mmwave}\xspace}
\newcommand{\mimo}{\gls{mimo}\xspace}

\newcommand{\rf}{\gls{rf}\xspace}

\newcommand{\fg}{\gls{5g}\xspace}

\newcommand{\iab}{\gls{iab}\xspace}

\newcommand{\adcs}{\glspl{adc}\xspace}

\newcommand{\bs}{\gls{bs}\xspace}

\newcommand{\gsnr}{\gls{snr}\xspace}
\newcommand{\ginr}{\gls{inr}\xspace}

\newcommand{\gpsnr}{\glspl{snr}\xspace}
\newcommand{\gpinr}{\glspl{inr}\xspace}
\newcommand{\gpsinr}{\glspl{sinr}\xspace}

\newcommand{\tdd}{\gls{tdd}\xspace}

\newcommand{\secref}[1]{Section~\ref{#1}}

\newcommand{\figref}[1]{\figurename~\ref{#1}}
\newcommand{\algref}[1]{Algorithm~\ref{#1}}

\newcommand{\thmref}[1]{Theorem~\ref{#1}}

\usepackage{mathtools}

\begin{document}
	
%
\title{\textsc{LoneSTAR}: Analog Beamforming Codebooks for Full-Duplex Millimeter Wave Systems}

%
%
%

\author{%
	Ian~P.~Roberts,~%
    Sriram Vishwanath,~%
	and Jeffrey~G.~Andrews%
    \thanks{The authors are with the 6G@UT Research Center and the Wireless Networking and Communications Group at the University of Texas at Austin. This work is an extension of \cite{roberts_2021_robustcb}. Corresponding author: I.~P.~Roberts (ipr@utexas.edu).} 
}

\maketitle





\begin{abstract}
    \edit{
This work develops \lonestar, a novel enabler of full-duplex \mmwave communication systems through the design of analog beamforming codebooks.
\lonestar codebooks deliver high beamforming gain and broad coverage while simultaneously reducing the self-interference coupled by transmit and receive beams at a full-duplex \mmwave transceiver.
Our design framework accomplishes this by tolerating some variability in transmit and receive beamforming gain to strategically shape beams that reject self-interference spatially while accounting for digitally-controlled analog beamforming networks and self-interference channel estimation error.
By leveraging the coherence time of the self-interference channel, a \mmwave system can use the same \lonestar design over many time slots to serve several downlink-uplink user pairs in a full-duplex fashion without the need for additional self-interference cancellation.
Compared to those using conventional codebooks, full-duplex \mmwave systems employing \lonestar codebooks can mitigate higher levels of self-interference, tolerate more cross-link interference, and demand lower SNRs in order to outperform half-duplex operation---all while supporting beam alignment.
This makes \lonestar a potential {standalone} solution for enabling {simultaneous transmission and reception} in \mmwave systems, from which it derives its name.
}

\end{abstract}




\glsresetall


\section{Introduction} \label{sec:introduction}

Codebook-based analog beamforming is a critical component of \mmwave communication systems \cite{heath_overview_2016,ethan_beam}.
Rather than measure a high-dimensional \mimo channel and subsequently configure an analog beamforming network, modern \mmwave systems instead rely on beam alignment procedures to identify promising beamforming directions, typically via exploration of a codebook of candidate beams \cite{heath_overview_2016,ethan_beam,junyi_wang_beam_2009}.
This offers a simple and robust way to configure an analog beamforming network without downlink/uplink \mimo channel knowledge \textit{a priori}, which is expensive to obtain in practice. 
In conventional \mmwave systems, it is relatively straightforward to design a beamforming codebook that delivers coverage over a desired region \cite{junyi_wang_beam_2009}.
For instance, conjugate beamforming and \gls{dft} beamforming are simple ways to construct a set of beams that steer high gain in specific directions using digitally-controlled phase shifters \cite{heath_lozano}.
Other codebook designs make use of amplitude control via digitally-controlled attenuators or \glspl{vga} to shape beams with suppressed side lobes in an effort to reduce interference \cite{phased_array_handbook,taylor_1955,taylor_SAR_1995}.


The design of beamforming codebooks for in-band full-duplex \mmwave communication systems, on the other hand, is not so straightforward.
A full-duplex \mmwave transceiver with separate transmit and receive arrays, for instance, will juggle a transmit beam and a receive beam at the same time over the same frequency spectrum, potentially serving two different half-duplex users.
These transmit and receive beams couple over the \mimo channel that manifests between the arrays, plaguing a desired receive signal with self-interference.
In a recent measurement campaign \cite{roberts_att_angular}, we observed that the degree of this self-interference is typically prohibitively high for full-duplex operation in practical \mmwave systems.
This, along with the fact that modern \mmwave systems rely on codebook-based beam alignment, motivates the need for beamforming codebooks designed specifically for full-duplex \mmwave systems---as opposed to using conventional codebooks made for half-duplex systems \cite{roberts_wcm}. 
In this work, we present \lonestar, the first framework for designing analog beamforming codebooks for full-duplex \mmwave communication systems. 
\lonestar produces transmit and receive codebooks that deliver high beamforming gain and provide broad coverage to downlink and uplink users while coupling low self-interference between the transmit and receive beams of a full-duplex transceiver.
\edit{
    By supporting beam alignment and reducing self-interference to levels near or below the noise floor regardless of which transmit and receive beams are selected from its codebooks, \lonestar can serve as a potential standa\underline{lone} solution for \underline{s}imultaneous \underline{t}ransmission \underline{a}nd \underline{r}eception, from which it derives its name.
}
Rather than using analog and digital self-interference cancellation to enable full-duplex as was commonly done in sub-6 GHz systems \cite{ashu_inband_jsac_2014}, recent work has proposed solely using beamforming to mitigate self-interference in \mmwave communication systems \cite{roberts_wcm,satyanarayana_hybrid_2019,liu_beamforming_2016,lopez_prelcic_2019_analog,lopez-valcarce_beamformer_2019,prelcic_2019_hybrid,zhu_uav_joint_2020,cai_robust_2019,koc_ojcoms_2021,lopez_analog_2022,da_silva_2020,roberts_bflrdr,roberts_steer}.
Such methods largely aim to take advantage of the high-dimensional spatial domain afforded by dense \mmwave antenna arrays to tackle self-interference via strategic design of transmit and receive beamformers.
However, there are key practical hurdles neglected by existing designs that we address in this work holistically.
Some designs \cite{liu_beamforming_2016,satyanarayana_hybrid_2019,prelcic_2019_hybrid,lopez_prelcic_2019_analog,lopez-valcarce_beamformer_2019,zhu_uav_joint_2020,cai_robust_2019,koc_ojcoms_2021,lopez_analog_2022,da_silva_2020} have neglected codebook-based analog beamforming, assuming the ability to fine-tune each phase shifter in real-time, and do not account for digitally-controlled phase shifters of practical analog beamforming networks.
Most designs have assumed a lack of amplitude control, even though it is not uncommon to have both digitally-controlled phase shifters and attenuators or \glspl{vga} in practice \cite{array_60GHz_2011,ibm_array_2017,anokiwave_5G}. 
In addition, existing solutions \cite{liu_beamforming_2016,satyanarayana_hybrid_2019,prelcic_2019_hybrid,lopez_prelcic_2019_analog,lopez-valcarce_beamformer_2019,zhu_uav_joint_2020,cai_robust_2019} have assumed real-time knowledge of the downlink and uplink \mimo channels---along with that of the self-interference \mimo channel---to configure transmit and receive beamforming. 
As such, they neglect beam alignment and rely on frequent estimation of high-dimensional \mimo channels, both of which are practical shortcomings.
Furthermore, the majority of existing designs are for hybrid digital/analog beamforming architectures and are not necessarily suitable for analog-only beamforming systems, which are common in practice.






To our knowledge, the only existing beamforming-based full-duplex solution for analog beamforming \mmwave systems that supports codebook-based beam alignment is our recent work called \steer \cite{roberts_steer}.
The work herein and that of \steer tackle a very similar problem but their approaches differ significantly due to differences in the assumption of self-interference channel knowledge.
In \steer, we relied on measurements of candidate beam pairs to strategically select transmit and receive beams that couple low self-interference while maintaining high gain in desired directions. 
In this work, rather than use measurements of self-interference, we instead rely on (imperfect) estimation of the self-interference \mimo channel, which \lonestar uses to construct transmit and receive codebooks such that every transmit-receive beam pair couples low self-interference.
\steer was validated using an actual 28 GHz phased array platform and proved to be a promising full-duplex solution, but it remains unclear if \steer's success will translate to other platforms. 
\edit{
Presented herein, \lonestar is more flexible to the particular self-interference channel than \steer by making use of more degrees of freedom, meaning it has the potential to better generalize across \mmwave platforms.
}


\textbf{\lonestar: Beamforming codebooks for full-duplex \mmwave systems.}
\edit{In this work, we present \lonestar, a novel approach to enable full-duplex \mmwave systems through the use of analog beamforming codebooks strategically designed to reduce self-interference and simultaneously deliver high beamforming gain.
We present an optimization framework for designing \lonestar codebooks that minimize self-interference coupled between all possible transmit-receive beam pairs while constraining the coverage they provide over defined service regions.
A full-duplex \mmwave transceiver can independently select any transmit and any receive beam from \lonestar codebooks and, with reasonable confidence, can expect low self-interference.
\lonestar codebooks are designed at the full-duplex transceiver following estimation of the self-interference \mimo channel (which need not be perfect), do not require downlink/uplink \mimo channel knowledge, and do not demand any over-the-air feedback to/from users.
After their construction, \lonestar codebooks can be used to conduct beam alignment and serve any downlink-uplink user pair in a full-duplex fashion thereafter with reduced self-interference. 
Importantly, \lonestar accounts for limited phase and amplitude control present in practical analog beamforming networks. 
}


\textbf{Validation of \lonestar as a full-duplex solution.}
\edit{
    We use extensive simulation to evaluate \lonestar, which highlights its potential to enable full-duplex operation in \mmwave systems without any additional self-interference cancellation.
We compare full-duplex performance with \lonestar codebooks against that with conventional codebooks to illustrate that \lonestar can supply an additional $10$--$50$ dB of robustness to self-interference, depending on hardware constraints, without sacrificing high beamforming gain.
This translates to impressive spectral efficiency gains that, with high-resolution phase shifters and attenuators, can approach the interference-free capacity of conventional codebooks.
Simulation further illustrates \lonestar's robustness to self-interference channel estimation error and to cross-link interference between downlink and uplink users. 
Our results motivate a variety of future work including self-interference channel modeling and reliable self-interference channel estimation.
}

\comment{
Codebook-based analog beamforming is a critical component of modern \mmwave systems \cite{heath_overview_2016}.
Rather than measure the complete over-the-air channel and subsequently configure analog beamformers, current \mmwave systems instead rely on beam alignment procedures to more quickly identify promising transmit and receive beamforming directions, typically via exploration of a codebook of candidate analog beams \cite{heath_overview_2016,junyi_wang_beam_2009}.
This offers a simple and robust way to configure dozens of phase shifters and attenuators without extensive channel knowledge \textit{a priori}.

Designing analog beamforming codebooks that span a desired coverage region is relatively straightforward for conventional half-duplex \mmwave systems \cite{junyi_wang_beam_2009}.
Conjugate beamforming (i.e., matched filter beamforming), for example, is a simple way to construct a set of beams that serve desired directions and only requires phase shifters, not attenuators \cite{heath_lozano}.
Other designs leverage attenuators to shape beams that exhibit wider main lobes and suppress side lobes, for example, which can reduce beam misalignment losses and inter-user interference.

Equipping \mmwave systems with full-duplex capability is an attractive proposition at not only the physical layer but also as a deployment solution through \iab \cite{roberts_wcm}.
To enable \mmwave full-duplex, recent work \cite{satyanarayana_hybrid_2019,liu_beamforming_2016,lopez_prelcic_2019_analog,lopez-valcarce_beamformer_2019,prelcic_2019_hybrid,zhu_uav_joint_2020,cai_robust_2019,roberts_2019_bfc,roberts_equipping_2020,roberts_bflrdr} has explored beamforming-based self-interference mitigation in various contexts.
Some designs \cite{liu_beamforming_2016,satyanarayana_hybrid_2019,prelcic_2019_hybrid,lopez_prelcic_2019_analog,lopez-valcarce_beamformer_2019,zhu_uav_joint_2020,cai_robust_2019} have neglected codebook-based analog beamforming, assuming the ability to fine-tune each phase shifter dynamically, and do not account for \textit{digitally}-controlled phase shifters.
Most designs have assumed a lack of amplitude control, even though it is not uncommon to have both phase shifters and attenuators in practical analog beamforming networks.
In addition, existing solutions \cite{liu_beamforming_2016,satyanarayana_hybrid_2019,prelcic_2019_hybrid,lopez_prelcic_2019_analog,lopez-valcarce_beamformer_2019,zhu_uav_joint_2020,cai_robust_2019,roberts_2019_bfc} have assumed transmit and receive channel knowledge along with that of the self-interference channel to configure precoding and combining---meaning they neglect beam alignment schemes and rely on highly dynamic updates as the transmit and receive channels change.

Like half-duplex \mmwave systems, one with full-duplex capability will presumably conduct codebook-based beam alignment on its transmit link \textit{and} receive link, meaning it will juggle a transmit beam and receive beam concurrently, which couple together via the self-interference channel.
Off-the-shelf analog beamforming codebooks that were designed for half-duplex settings may be undesirable in full-duplex settings since they do not necessarily offer robustness to self-interference \cite{roberts_wcm}.
Instead, we design analog beamforming codebooks for full-duplex that reliably deliver high beamforming gain to users and simultaneously reject self-interference regardless of which transmit and receive beams are used.  Given such a codebook, standard beam alignment procedures that are self-interference agnostic can be utilized in a full-duplex system.
This is a desirable practical outcome from our approach. 

Among existing literature, we are not aware of any work on the design of analog beamforming \textit{codebooks} for \mmwave full-duplex.  Our main contribution is a methodology for designing transmit and receive analog beamforming codebooks that reduces the average self-interference coupled between transmit-receive beam pairs while also guaranteeing the beamforming gain they provide. We present an algorithm for approximately solving for our design that addresses the non-convexity posed by digitally-controlled phase shifters and attenuators.
Results indicate that our design offers $20$--$50$ dB of added robustness to self-interference by strategically shaping beams with gain comparable to conventional codebooks. 

As a result, codebooks designed with our approach could improve existing \mmwave full-duplex work \cite{roberts_2019_bfc,cai_robust_2019,roberts_equipping_2020,roberts_bflrdr} that accounts for codebook-based analog beamforming.  Hybrid digital/analog beamforming full-duplex systems could leverage our codebooks by weakening the \textit{effective} self-interference channel post beam alignment. Perhaps most excitingly, a \mmwave system employing our codebook design can in principle execute beam alignment and then seamlessly operate in a full-duplex fashion, thanks to codebooks whose beams offer inherent robustness to self-interference.  
}

\comment{
Codebook-based analog beamforming is a critical component to the deployment of \mmwave systems.
Rather than measure the over-the-air channel and subsequently configure analog beamformers, practical \mmwave systems typically employ beam alignment to identify directions where energy should be steered to close the link between two devices.
The analog beamforming network---which is comprised of many digitally controlled phase shifters and possibly attenuators---is configured to steer in promising directions via a lookup from a codebook of beams.
This codebook-based beam alignment offers a simple and robust approach to analog beamforming versus the system setting each of the dozens (potentially hundreds) of phase shifters and attenuators one-by-one.

Designing useful analog beamforming codebooks is relatively straightforward for half-duplex systems.
Given the phase shifter and attenuator resolution, number of antennas, and desired steering resolution, a set of highly directional beams that span a desired coverage region can be fairly easily designed.
Conjugate beamforming, for example, is a simple way to construct a set of beams that serve desired directions and only requires phase shifters, not attenuators.
With sufficient phase resolution, conjugate beams can provide full array gain in any desired direction.
Other designs leverage attenuators to shape beams that, for example, yield wider main lobes and suppress side lobes, which can help avoid loss due to beam misalignment and reduce interference, respectively.

In \mmwave systems that full-duplex transmission and reception, such as full-duplex \iab proposed in 3GPP BLANK, using analog beamforming codebooks that were designed for half-duplex settings may be highly undesirable, however.
This is because, at a full-duplex transceiver, the transmit and receive beams couple together via a self-interference channel. 
Off-the-shelf analog beamforming codebooks used for half-duplex do not necessarily offer resilience to rejecting this self-interference---after all, they weren't designed for such.
This motivates us to design analog beamforming codebooks for full-duplex systems to reduce the degree of self-interference coupled by all transmit-receive beam pairs.
By doing so, full-duplexing transmission and reception can be facilitated in analog-only and hybrid digital/analog beamforming \mmwave systems.

To design an analog beamforming codebook for \mmwave is difficult for a variety of reasons.
First and foremost, there has yet to be a verified, reliable self-interference channel model for \mmwave systems.
Even with a model, it will likely be difficult to create a codebook that is not system-dependent, given that each system will have a unique setup, infrastructure, and environment.
Together, these two facts make it difficult to create a universal and practical codebook.
While it is theoretically possible to estimate the self-interference channel at a system and subsequently design an analog beamforming codebook to reject such, this is not especially practical for a variety of reasons.
Given that self-interference channel is observed through the transmit and receive beamformers---i.e., \dacs and \adcs are not per-antenna---it can be difficult to extract details about the over-the-air channel reliably. 
This can be attributed to the artifacts that may be inherent to the array's elements, beamforming weights, power amplifiers, phase inconsistencies, miscalibration, and the like.

%
%
%
%
%
%
%

%
%
}



\section{System Model} \label{sec:system-model}

\begin{figure}
    \centering
    \includegraphics[width=\linewidth,height=0.3\textheight,keepaspectratio]{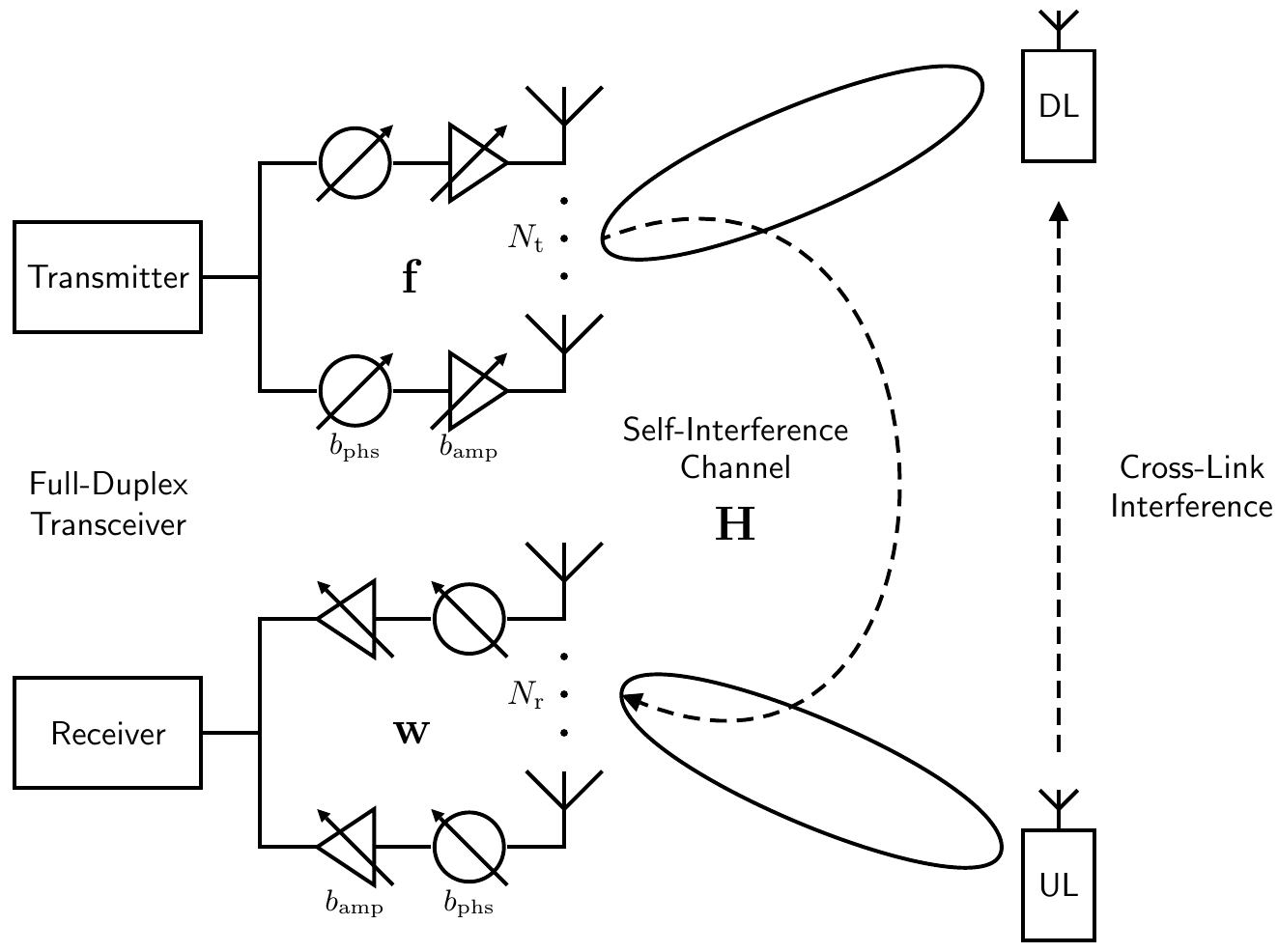}
    \caption{A full-duplex \mmwave \bs transmits to a downlink user while receiving from an uplink user in-band. In doing so, self-interference couples between its transmit beam $\vf$ and receive beam $\vw$ across the channel $\mH$, and cross-link interference is inflicted on the downlink user by the uplink user.}
    \label{fig:system}
\end{figure}

Illustrated in \figref{fig:system}, we consider a \mmwave \bs that employs separate, independently-controlled antenna arrays for transmission and reception that are used to serve a downlink user and uplink user in a full-duplex fashion (simultaneously and in-band).
One may consider full-duplex \iab, for instance, as an attractive application of this work in \mmwave cellular networks \cite{iab,3GPP_IAB_2,gupta_fdiab_arxiv,suk_iab_2022}.
Let $\Nt$ and $\Nr$ be the number of transmit and receive antennas, respectively, at the arrays of the full-duplex \bs.
We denote as $\atx{\theta} \in \setvectorcomplex{\Nt}$ and $\arx{\theta} \in \setvectorcomplex{\Nr}$ its transmit and receive array response vectors, respectively, in the direction $\theta$. 
These array response vectors for any $\theta$ can be computed based on the \bs array geometries \cite{heath_lozano,balanis}, and we follow the convention $\normtwo{\atx{\theta}}^2 = \Nt$ and $\normtwo{\arx{\theta}}^2 = \Nr$.
To simplify presentation, we assume users are single-antenna devices, but this is not a necessary assumption. 

We consider an analog beamforming system at the \bs, though the codebook design herein could also be used by hybrid beamforming systems.
Let $\vf \in \setvectorcomplex{\Nt}$ be the beamforming vector used at the transmitter of the \bs and $\vw \in \setvectorcomplex{\Nr}$ be the beamforming vector used at its receiver.
We consider analog beamforming networks equipped with phase shifters and attenuators, which are digitally controlled with $\bitsphase$ and $\bitsamp$ bits, respectively.\footnote{Existing \mmwave literature often considers networks with only phase shifters, but it is not uncommon for practical \mmwave arrays to also be equipped with attenuators or \glspl{vga} for amplitude control \cite{array_60GHz_2011,ibm_array_2017,anokiwave_5G}. Our formulation could also capture beamforming networks employing \glspl{vga} (rather than attenuators) by including their maximum gain settings in \eqref{eq:unit-f-w}, for instance.}
Each beamforming weight must not exceed unit magnitude, understood by the fact that attenuators are used for amplitude control.
\begin{gather}
\bars{\entry{\vf}{m}} \leq 1 \ \forall \ m = 1, \dots, \Nt, \qquad \bars{\entry{\vw}{n}} \leq 1 \ \forall \ n = 1, \dots, \Nr \label{eq:unit-f-w}
\end{gather}
Directly from \eqref{eq:unit-f-w}, we have $\normtwo{\vf}^2 \leq \Nt$ and $\normtwo{\vw}^2 \leq \Nr$.
Capturing limited phase and amplitude control, let $\precb \subset \setvectorcomplex{\Nt}$ and $\comcb \subset \setvectorcomplex{\Nr}$ be the sets of all possible physically realizable beamforming vectors $\vf$ and $\vw$, respectively; hence, it is practically required that $\vf \in \precb$ and $\vw \in \comcb$.
We assume the \bs and uplink user transmit with powers $\powertxbs$ and $\powertxue$, respectively. 
Additive noise powers per antenna at the \bs and downlink user are respectively $\powernoisebs$ and $\powernoiseue$. 


Taking the perspective of the full-duplex \bs, let $\vhtx\ctrans \in \setmatrixcomplex{1}{\Nt}$ be the transmit channel vector from the \bs to the downlink user.
Let $\vhrx \in \setmatrixcomplex{\Nr}{1}$ be the receive channel from the uplink user to the \bs.
We normalize each of these channels such that $\normtwo{\vhtx}^2 = \Nt$ and $\normtwo{\vhrx}^2 = \Nr$ and abstract out their large-scale gains as inverse path losses $\Gtx^2$ and $\Grx^2$, respectively.
Since transmission and reception happen simultaneously and in-band, a self-interference channel $\mH \in \setmatrixcomplex{\Nr}{\Nt}$ manifests between the transmit and receive arrays of the full-duplex \bs.
Currently, there is not a reliable, measurement-backed model for the self-interference channel $\mH$ \cite{roberts_wcm,roberts_att_angular}.
As such, the design herein is not based on any assumption of $\mH$, and we evaluate our design using multiple channel models in \secref{sec:simulation-results}.
To abstract out the large-scale gain of the self-interference channel from its spatial characteristics, we impose $\normfro{\mH}^2 = \Nt \cdot \Nr$
and let $G^2$ be its inverse path loss based on the inherent isolation between the arrays (practically, $G^2 \ll 0$ dB).
With the downlink and uplink users operating simultaneously and in-band, a scalar cross-link interference channel $\hcl$ arises between them.
We normalize $\bars{\hcl}^2 = 1$ and let $\Gcl^2$ be the inverse path loss of the cross-link interference channel, which practically depends on the users' locations and the environment.




\section{Problem Formulation and Motivation} \label{sec:problem-formulation}

Using the established system model, we now further formulate and motivate this work.
Again taking the perspective of the full-duplex \bs, we refer to downlink as the \textit{transmit link} and to uplink as the \textit{receive link}.
For some transmit and receive beams $\vf$ and $\vw$, the \glspl{snr} of the transmit and receive links are respectively
\begin{align}
\snrtx &= \frac{\powertxbs \cdot \Gtx^2 \cdot \bars{\vhtx\ctrans \vf}^2}{\Nt \cdot \powernoiseue}, \qquad 
\snrrx = \frac{\powertxue \cdot \Grx^2 \cdot \bars{\vw\ctrans \vhrx}^2}{\normtwo{\vw}^2 \cdot \powernoisebs} \label{eq:snr-trx}
\end{align}
where dividing by $\Nt$ in $\snrtx$ handles power splitting within the transmit array and by $\normtwo{\vw}^{2}$ in $\snrrx$ handles noise combining at the receive array.
By virtue of the fact that 
$N^2 = \normtwo{\vh}^2 \cdot \normtwo{\vx}^2 = \max_{\vx} \ \bars{\vh\ctrans \vx}^2 \ \st \normtwo{\vx}^2 \leq N, \normtwo{\vh}^2 = N$,
the maximum \gpsnr are denoted with an overline as
\begin{align}
\snrtxbar &= \frac{\powertxbs \cdot \Gtx^2 \cdot \Nt}{\powernoiseue} \geq \snrtx, \qquad
\snrrxbar = \frac{\powertxue \cdot \Grx^2 \cdot \Nr}{\powernoisebs} \geq \snrrx.
\end{align}
In other words, these are achieved only when users are delivered maximum beamforming gain.


When juggling a transmit beam $\vf$ and receive beam $\vw$ in a full-duplex fashion, the \bs couples self-interference, leading to a receive link \ginr of
\begin{align}
\inrrx = \frac{\powertxbs \cdot G^2 \cdot \bars{\vw\ctrans \mH \vf}^2}{\Nt \cdot \normtwo{\vw}^2 \cdot \powernoisebs}
\end{align}
where $\bars{\vw\ctrans \mH \vf}^2$ captures the transmit and receive beam coupling over the self-interference channel.
Strategically steering $\vf$ and $\vw$ according to $\mH$ can therefore reduce self-interference.
Based on our normalizations, the worst-case coupling between transmit and receive beams at the full-duplex \bs is $\Nt^2 \cdot \Nr^2 = \max \bars{\vw\ctrans \mH \vf}^2$, 
leading to a maximum possible $\inrrx$ of
\begin{align}
\inrrxbar = \frac{\powertxbs \cdot G^2 \cdot \Nt \cdot \Nr}{\powernoisebs} \geq \inrrx
\end{align}
which we use as a metric to capture the inherent strength of self-interference at the \bs.
Note that $\inrrxbar$ depends solely on system parameters and artifacts at the \bs. 

We denote as $\inrtx$ the \ginr incurred at the downlink user due to cross-link interference.
\begin{align}
\inrtx = \frac{\powertxue \cdot \Gcl^2 \cdot \bars{\hcl}^2}{\powernoiseue}
\end{align}
Note that cross-link interference $\inrtx$ only depends on its inherent channel and path loss, not on $\vf$ nor $\vw$.
With all of this, the \gpsinr of the links are
\begin{align}
\sinrtx &= \frac{\snrtx}{1+\inrtx}, \qquad 
\sinrrx = \frac{\snrrx}{1+\inrrx}.
\end{align}
The achievable spectral efficiencies on each link, treating interference as noise, are
\begin{align}
\setx &= \logtwo{1 + \sinrtx}, \qquad
\serx = \logtwo{1 + \sinrrx}
\end{align}
whereas the capacities of the two links are $\captx = \logtwo{1 + \snrtxbar}$ and $\caprx = \logtwo{1 + \snrrxbar}$.

We assume no forms of active self-interference cancellation are employed at the \bs.
Instead, we will rely solely on beamforming to mitigate self-interference and enable full-duplex operation. 
The degree of self-interference coupled at the full-duplex \bs depends on the transmit beam $\vf$ and receive beam $\vw$, each of which also dictates its respective link's \gsnr.
As such, it is important that $\vf$ and $\vw$ couple low self-interference and deliver high $\snrtx$ and $\snrrx$ in an effort to achieve high sum spectral efficiency $\setx + \serx$.

In this work, we rely on estimation of the self-interference \mimo channel $\mH$ at the full-duplex \bs.
We do not assume perfect estimation of $\mH$ but rather model its imperfections as
\begin{align}
\mH = \mHbar + \mDelta, \qquad \entry{\mDelta}{i,j} \sim \distcgauss{0}{\epsilon^2} \ \forall \ i,j.  \label{eq:channel}
\end{align}
where $\mHbar$ is the estimate of $\mH$ and $\mDelta$ captures \iid Gaussian estimation error with mean $0$ and variance $\epsilon^2$. 
Self-interference channel estimation in \mmwave systems is an open research problem \cite{roberts_wcm}.
As such, the choice of a particular estimation error model is currently difficult to practically justify, but we remark that other error models (e.g., norm-based models) may be encompassed by our Gaussian model with arbitrarily high probability for appropriately chosen $\eps^2$.
It is our hope that $\mH$ can be estimated with high fidelity for three reasons.
First, the self-interference channel is likely much stronger than traditional downlink/uplink \mimo channels, meaning it may be estimated with higher accuracy.
Second, the estimation of $\mH$ takes place across the arrays of the \bs, meaning it does not consume resources for feedback and does not suffer from feedback quantization.
Third, we expect that the self-interference channel will be fairly static on the timescale of data transmissions, especially the portion caused by direct coupling between the arrays\footnote{In \cite{roberts_att_angular}, we observed that 28 GHz self-interference in an anechoic chamber was nearly static on the order of minutes, at least.}, which will presumably dominate.
Together, all of this suggests that the estimation of $\mH$ may be more thorough and precise than that of traditional downlink/uplink \mimo channels.
Nonetheless, our design in the next section incorporates channel estimation error.
Self-interference channel estimation and its coherence time would be valuable directions for future work, especially in practical environments.
The key practical advantages of \lonestar rely on the self-interference channel being sufficiently static over many downlink/uplink time slots, which we assume herein.

\comment{



Rather than obtaining over-the-air channel knowledge between two devices in order to configure $\prerf$ and $\comrf$, practical systems like \fg cellular and IEEE 802.11ad/ay have turned to codebook-based beam alignment.
By taking this approach, promising analog beams (e.g., those that offer high \gls{snr}) are identified by searching through the codebook and measuring the power delivered to a device being served.
This results in a significant complexity reduction in designing $\prerf$ and $\comrf$ and simplifies channel estimation and initial access.
Clearly, searching through $\prerfcb$ or $\comrfcb$ exhaustively during beam alignment is highly impractical.
Therefore, it is much more practical to build significantly reduced codebooks $\prerfcbbar$ and $\comrfcbbar$ from which analog precoders and combiners are drawn.
With this being said, one can see that a codebook-based analog beamforming leaves very little room for optimization: the only flexibility the system has in configuring its analog beamformers is choosing from the entries in the codebook. \edit{Improve.}

Consider our full-duplex setting where the analog precoder couples with the analog combiner via the self-interference channel.
Some transmit-receive beam pairs will inherently afford more isolation than others.
To reduce the amount of self-interference reaching the receiver, it would be ideal if all beam pairs afforded high isolation.
It is unlikely, however, for this to be the case when using ``off-the-shelf'' codebooks---after all, they were not designed with full-duplex in mind.
One may be inclined to believe that the extremely narrow beams that have become ubiquitous in \mmwave would offer robustness to coupling because of their pencil-like shape.
This is, however, not necessarily the case because these extremely narrow beams are only ``extremely narrow'' in the far-field and not necessarily so in near-field.
This is evidenced by \figref{fig:nf-patterns}, where BLANK.
This motivates us to design an analog precoding codebook $\prerfcbbar$ and combining codebook $\comrfcbbar$ that together offer high isolation across all transmit-receive beam pairs to ensure that no matter which directions a device transmits and receives from, it sees a relatively low degree of coupled self-interference.

En route to reducing the amount of self-interference coupled by the analog precoding and combining codebooks, we must ensure that the beams within the codebooks remain useful in their ability to serve users.
Since \mmwave communication relies on high analog beamforming gains for sufficient link margin, beams are only useful if they offer high gain.
Additionally, since users may exist anywhere within a service region, a codebook of beams must collectively maintain coverage to the service region.
Conventional codebooks certainly satisfy this definition of ``usefulness'' by design as exhibited by \figref{fig:conventional-codebooks}, but do not offer much robustness to self-interference as exhibited by 

With $\bitsamp$ bits of amplitude control per attenuator and $\bitsphase$ bits of phase control per phase shifter, there are a total of $2^{\bitsamp + \bitsphase}$ different ways to configure a \textit{single} analog beamforming weight.
With $\Na$ antennas, there are $2^{\Na\cdot\parens{\bitsamp+\bitsphase}}$ different ways to configure an analog beamforming vector.
With $L$ \rf chains, there are $2^{L\cdot\Na\cdot\parens{\bitsamp+\bitsphase}}$ different ways to configure an analog beamforming matrix.
To design our codebook, clearly searching exhaustively through $\prerfcb$ or $\comrfcb$ for beams that supply good coverage with reduced self-interference coupling is impractical.
This motivates us to develop a systematic methodology to efficiently build our codebook, which we present in \secref{sec:contribution-i}.

\subsection{3-D Geometry}
Let $\dirset$ be the set of all azimuth-elevation pairs $(\theta,\phi)$ where $\theta \in [0,2\pi]$ and $\phi \in [0,\pi]$.
\begin{align}
\dirset = \braces{(\theta,\phi) : \theta \in [0,2\pi], \phi \in [0,\pi]}
\end{align}
A vector in the direction $(\theta,\phi)$ is described by $\theta$ being angle between the positive $y$ axis and the vector's orthogonal projection onto the $x$-$y$ plane and $\phi$ being the angle from the $x$-$y$ plane to the vector itself.

\begin{definition}[Transmit and receive beam mappings]
	Let $f : \dirset \to \precb$ be a mapping from transmit direction to analog precoding weights $\vf$ as
	\begin{align}
	\vf = f(\thetatx,\phitx)
	\end{align}
	where $(\thetatx,\phitx) \in \dirset$ is a transmit direction.
	Similarly, let $w : \set{A} \to \comcb$ be a mapping from receive direction to analog combining weights $\vw$ as
	\begin{align}
	\vw = w(\thetarx,\phirx)
	\end{align}
	where $(\thetarx,\phirx) \in \dirset$ is a receive direction.
	The mappings $f$ and $w$ capture arbitrary beamforming design methodologies, which are implemented by system engineers or supplied by the array manufacturer.
\end{definition}
}

\section{Analog Beamforming Codebook Design for Full-Duplex mmWave Systems}
\label{sec:contribution}

Practical \mmwave systems---which have operated in a half-duplex fashion thus far---rely on beam alignment to identify beams that deliver high beamforming gain when serving a particular user.
This is typically done by sweeping a set of candidate beams called a \textit{codebook} and measuring the \gls{rsrp} for each beam candidate in order to select which is used to close a link. 
This procedure allows a \mmwave system to reliably deliver high beamforming gain without the need for downlink/uplink \mimo channel knowledge. 

As mentioned in the introduction, plenty of existing work has designed particular transmit and receive beams $\vf$ and $\vw$ to use at the \bs to reduce self-interference when full-duplexing downlink and uplink users, but most existing designs have a number of practical shortcomings.
In this work, we instead design transmit and receive beam codebooks $\precbbar$ and $\comcbbar$, from which the \bs can draw for $\vf$ and $\vw$ via conventional beam alignment.
The goal of our design---which we call \lonestar---is to create codebooks that a full-duplex \mmwave \bs can use to deliver high beamforming gain to users while simultaneously mitigating self-interference.
This will be achieved by strategically designing $\precbbar$ and $\comcbbar$ such that their transmit and receive beams couple low self-interference, regardless of which $\vf \in \precbbar$ and $\vw \in \comcbbar$ are selected following beam alignment.
In contrast, codebooks created for conventional \mmwave systems can deliver high beamforming gain but are not necessarily robust to self-interference if used in full-duplex systems \cite{roberts_wcm}. 
In fact, this was confirmed in our recent measurement campaign \cite{roberts_att_angular} and in our recent work \cite{roberts_steer}, both of which showed that conventional beams typically couple prohibitively high amounts of self-interference if used in a practical full-duplex \mmwave system.


\begin{figure*}
    \centering
    \includegraphics[width=\linewidth,height=\textheight,keepaspectratio]{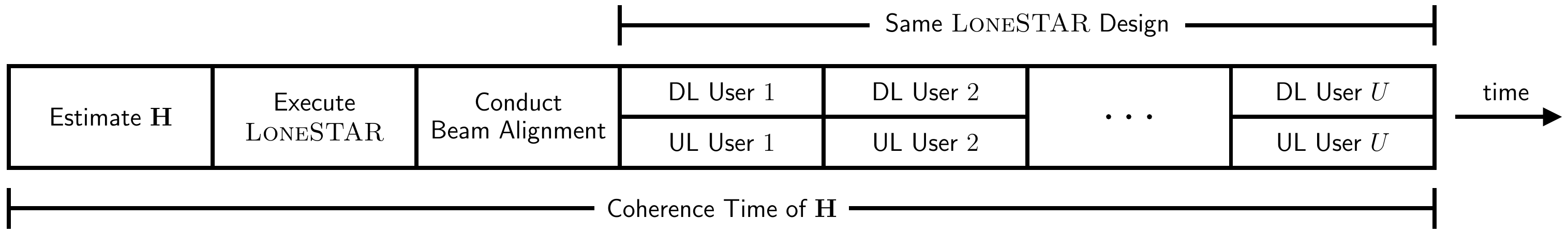}
    \caption{Following estimation of the self-interference channel $\mH$, transmit and receive beam codebooks designed with \lonestar can be used for beam alignment and subsequently to serve any downlink and uplink users in a full-duplex fashion.}
    \label{fig:timeline}
\end{figure*}

\begin{figure*}
    \centering
    \includegraphics[width=\linewidth,height=\textheight,keepaspectratio]{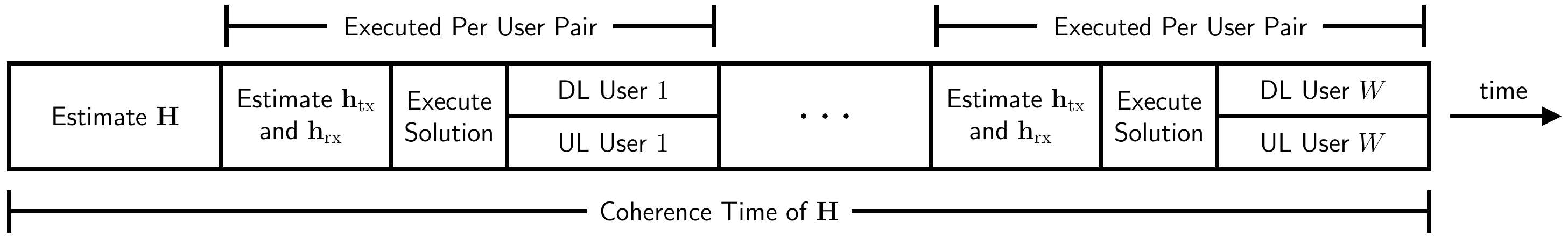}
    \caption{Most existing beamforming-based full-duplex \mmwave solutions require estimation of the downlink and uplink channels and execution of their design for each user pair, leading to a total overhead that scales with the number of users served.}
    \label{fig:timeline-2}
\end{figure*}

%

\edit{
Our envisioned use of \lonestar is depicted in \figref{fig:timeline}, which illustrates that our codebook design need only be executed once per coherence time of $\mH$, allowing the same full-duplex solution to be used over many downlink-uplink user pairs (i.e., data transmissions).
This is a key practical advantage of \lonestar versus most existing \mmwave full-duplex solutions, which we depict in \figref{fig:timeline-2}.
Most often, existing designs tailor their solutions uniquely for each downlink-uplink user pair and warrant estimation of the downlink and uplink \mimo channels, which itself is impractical in today's systems. 
On the other hand, a single \lonestar design can serve any sequence of downlink-uplink user pairs in a full-duplex fashion.
Moreover, existing designs typically introduce over-the-air feedback to and from users, whereas \lonestar does not require this---its design taking place completely at the full-duplex transceiver. 
Naturally, the coherence time of \mmwave self-interference, which is not currently well understood, will be an important factor in justifying the use of \lonestar since it dictates how frequently $\mH$ must be estimated and how often \lonestar needs to be re-executed.
It should be noted, however, that most existing designs also rely on estimation of the self-interference channel, meaning they too are subjected to its coherence time.
If our design sufficiently mitigates self-interference between all beam pairs, a \mmwave \bs employing \lonestar codebooks can seamlessly operate in a full-duplex fashion without any additional self-interference cancellation---any transmit and receive beams selected via beam alignment will couple low self-interference.
}

\subsection{Quantifying Our Codebook Design Criteria}
\begin{figure}
    \centering
    \includegraphics[width=\linewidth,height=0.25\textheight,keepaspectratio]{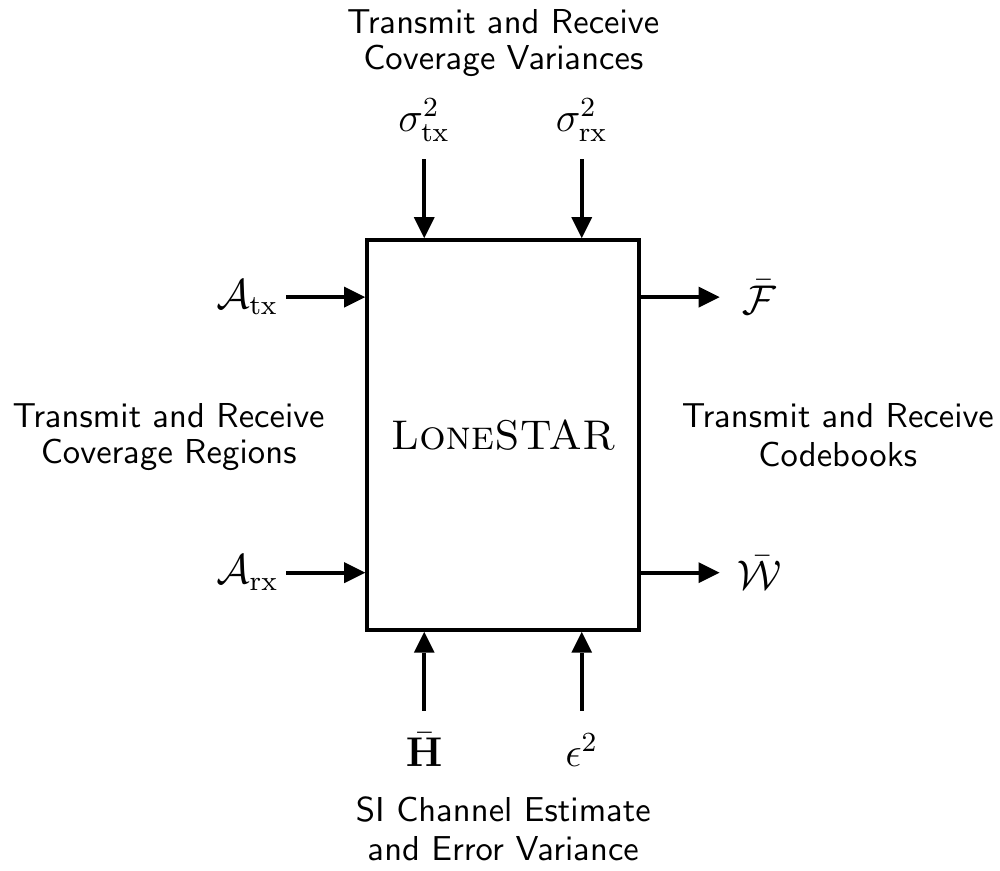}
    \caption{Using an estimate of the self-interference channel and knowledge of the estimation error variance, \lonestar produces transmit and receive beamforming codebooks that serve some coverage regions with quality based on tolerated coverage variances.}
    \label{fig:block}
\end{figure}

In pursuit of transmit and receive beam codebooks that reliably offer high sum spectral efficiency $\setx + \serx$, we desire sets of transmit and receive beams that can deliver high $\snrtx$ and $\snrrx$ while coupling low self-interference $\inrrx$.
As mentioned, cross-link interference $\inrtx$ is fixed for a given downlink-uplink user pair since it does not depend on the choice of $\vf$ nor $\vw$.
In the remainder of this section, we first quantify our codebook design criteria and then assemble a design problem that tackles both objectives jointly.
As we proceed with outlining our design, please refer to \figref{fig:block} for an illustration of the inputs and outputs of \lonestar.

\begin{definition}[Transmit and receive coverage regions]
Let $\dirtxset$ be a set of $\Mtx$ directions---or a \textit{transmit coverage region}---that our transmit codebook $\precbbar$ aims to serve, and let $\dirrxset$ be a \textit{receive coverage region} defined analogously.
\begin{align}
\dirtxset = \braces{\dirtx\idx{i} : i = 1, \dots, \Mtx}, \qquad \dirrxset = \braces{\dirrx\idx{j} : j = 1, \dots, \Mrx}
\end{align}
Here, $\dirtx\idx{i}$ and $\dirrx\idx{j}$ each may capture azimuthal and elevational components of a given direction.
\end{definition}


\begin{definition}[Transmit and receive beamforming codebooks] \label{def:codebooks}
We index the $\Mtx$ beams in $\precbbar$ and the $\Mrx$ beams in $\comcbbar$ according to $\dirtxset$ and $\dirrxset$, respectively, as $\precbbar = \braces{\vf_1, \dots, \vf_{\Mtx}}$ and $\comcbbar = \braces{\vw_1, \dots, \vw_{\Mrx}}$, 
where $\vf_i \in \precb$ is responsible for transmitting toward $\dirtx\idx{i}$ and $\vw_j \in \comcb$ is responsible for receiving from $\dirrx\idx{j}$. 
Using this notation, we build the transmit codebook matrix $\precbmat$ by stacking transmit beams $\vf_i$ as follows and the receive codebook matrix $\comcbmat$ analogously.
\begin{align}
\precbmat = 
\begin{bmatrix}
\vf_1 & \vf_2 & \cdots & \vf_{\Mtx}
\end{bmatrix} \in \setmatrixcomplex{\Nt}{\Mtx}, \qquad
\comcbmat = 
\begin{bmatrix}
\vw_1 & \vw_2 & \cdots & \vw_{\Mrx}
\end{bmatrix} \in \setmatrixcomplex{\Nr}{\Mrx} \label{eq:unpack-2}
\end{align}
\end{definition}

\begin{definition}[Transmit and receive beamforming gains]
    Let $\Gtx^2\parens{\dirtx\idx{i}}$ be the transmit beamforming gain afforded by $\vf_i$ in the direction $\dirtx\idx{i}$ and $\Grx^2\parens{\dirrx\idx{j}}$ be the receive beamforming gain afforded by $\vw_j$ in the direction $\dirrx\idx{j}$, expressed as
    \begin{align}
    \Gtx^2\parens{\dirtx\idx{i}} &= \bars{\atx{\dirtx\idx{i}}\ctrans  \vf_i}^2 \leq \Nt^2 \label{eq:Gtx}, \qquad 
    \Grx^2\parens{\dirrx\idx{j}} = \bars{\arx{\dirrx\idx{j}}\ctrans  \vw_j}^2 \leq \Nr^2. 
    \end{align}
    The transmit and receive beamforming gains are upper-bounded by $\Nt^2$ and $\Nr^2$, respectively, by virtue of normalizations presented in our system model. 
    Maximum beamforming gains can be achieved toward any direction $\dir$ with beams $\vf = \atx{\dir}$ and $\vw = \arx{\dir}$, respectively, often referred to as \textit{conjugate beamforming} or \textit{matched filter beamforming} \cite{heath_lozano}.
\end{definition}

\begin{definition}[Average self-interference coupled]
The coupling between the $i$-th transmit beam $\vf_i$ and the $j$-th receive beam $\vw_j$ is captured by the product $\vw_j\ctrans \mH \vf_i$, which can be extended to all transmit-receive beam pairs as $\comcbmat\ctrans \mH \precbmat \in \setmatrixcomplex{\Mrx}{\Mtx}$.
\lonestar will aim to minimize the coupling across all beam pairs and does so by minimizing the average $\inrrx$ coupled between beam pairs, which can be written as
\begin{align}
\inrrxavg 
&= \frac{1}{\Mtx \cdot \Mrx} \cdot \sum_{i=1}^{\Mtx} \sum_{j=1}^{\Mrx}  \frac{\powertxbs \cdot G^2 \cdot \bars{\vw_j\ctrans \mH \vf_i}^2}{\Nt \cdot \normtwo{\vw_j}^2 \cdot \powernoisebs} 
\approx \frac{\powertxbs \cdot G^2}{\Nt \cdot \Nr \cdot \powernoisebs} \cdot \frac{\normfro{\mW\ctrans \mH \mF}^2}{\Mtx \cdot \Mrx}. \label{eq:inr-rx-avg}
\end{align}
where the approximation becomes equality when $\normtwo{\vw_j}^2 = \Nr$ for all $j$.
Minimizing $\inrrxavg$ can be accomplished (approximately) by minimizing $\normfro{\mW\ctrans \mH \mF}^2$ through the design of $\mF$ and $\mW$ for a given $\mH$ since all other terms are constants.
In \lonestar's attempt to minimize $\inrrxavg$, it must also ensure that the beams in $\precbmat$ and $\comcbmat$ remain useful in serving users; of course, $\precbmat$ and $\comcbmat$ minimize $\inrrxavg$ when driven to $\zeromat$ if unconstrained.
We now quantify a means to constrain \lonestar to ensure adequate service is maintained in its aim to minimize $\inrrxavg$.
\end{definition}

\begin{definition}[Transmit and receive coverage variances]
To offer flexibility in minimizing $\inrrxavg$, \lonestar tolerates some sacrifices in beamforming gain over the transmit and receive coverage regions.
In other words, we allow for transmit and receive beamforming gains less than $\Nt^2$ and $\Nr^2$ over $\txdirsetcb$ and $\rxdirsetcb$ in pursuit of reduced self-interference. 
Let $\sigmatx^2 \geq 0$ be the maximum variance tolerated in delivering maximum transmit gain across $\txdirsetcb$, normalized to $\Nt^2$.
Defining $\sigmarx^2 \geq 0$ analogously, we express these tolerated coverage variances as 
\begin{align}
\frac{1}{\Mtx} \cdot \sum_{i=1}^{\Mtx} \frac{\bars{\Nt - \Gtx\parens{\dirtx\idx{i}}}^2}{\Nt^2} &\leq \sigmatx^2 \label{eq:sigma-tx}, \qquad \quad
\frac{1}{\Mrx} \cdot \sum_{j=1}^{\Mrx} \frac{\bars{\Nr - \Grx\parens{\dirrx\idx{j}}}^2}{\Nr^2} \leq \sigmarx^2. 
\end{align}
Here, $\sigmatx^2$ and $\sigmarx^2$ are design parameters that can be tuned by system engineers; increasing them will offer \lonestar greater flexibility in reducing $\inrrxavg$ but may degrade coverage since it permits more variability in beamforming gain over the service regions.
\end{definition}

\subsection{Assembling Our Codebook Design Problem}
In the previous subsection, we outlined that \lonestar will attempt to minimize the average self-interference coupled by transmit and receive beams while constraining the variance of beamforming gain delivered to specified transmit and receive coverage regions.
With expressions for our codebook design criteria in hand, we now turn our attention to assembling a formal design problem.
Let $\Atx \in \setmatrixcomplex{\Nt}{\Mtx}$ be the matrix of transmit array response vectors evaluated at the directions in $\dirtxset$, and let $\Arx \in \setmatrixcomplex{\Nr}{\Mrx}$ be defined analogously. 
\begin{align}
\Atx &= 
\begin{bmatrix}
\atx{\dirtx\idx{1}} & \atx{\dirtx\idx{2}} & \cdots & \atx{\dirtx\idx{\Mtx}}
\end{bmatrix} \\
\Arx &= 
\begin{bmatrix}
\arx{\dirrx\idx{1}} & \arx{\dirrx\idx{2}} & \cdots & \arx{\dirrx\idx{\Mrx}}
\end{bmatrix}
\end{align}
Recall, these can be computed directly based on the transmit and receive array geometries \cite{heath_lozano,balanis}.
Using these, the expressions of \eqref{eq:sigma-tx} can be written equivalently (up to arbitrary phase shifts of $\vf_i$ and $\vw_j$) as follows, which we refer to as \textit{coverage constraints}.
\begin{align}
\normtwo{\Nt\cdot \vone - \diag{\Atx\ctrans \precbmat}}^2 &\leq \sigmatx^2 \cdot \Nt^2 \cdot \Mtx \label{eq:coverage-constraint-tx} \\
\normtwo{\Nr\cdot \vone - \diag{\Arx\ctrans \comcbmat}}^2 &\leq \sigmarx^2 \cdot \Nr^2 \cdot \Mrx \label{eq:coverage-constraint-rx}
\end{align}
Note that the $(i,i)$-th entry of $\Atx\ctrans \precbmat$ has magnitude $\Gtx\parens{\dirtx\idx{i}}$ and the $(j,j)$-th entry of $\Arx\ctrans \comcbmat$ has magnitude $\Grx\parens{\dirrx\idx{j}}$.
By satisfying \eqref{eq:coverage-constraint-tx} and \eqref{eq:coverage-constraint-rx}, engineers can ensure that \lonestar codebooks adequately serve their coverage regions for appropriately chosen $\sigmatx^2$ and $\sigmarx^2$.


With coverage constraints \eqref{eq:coverage-constraint-tx} and \eqref{eq:coverage-constraint-rx}, we now formulate an analog beamforming codebook design in problem \eqref{eq:problem-desired-i}.
The goal of \lonestar is to design transmit and receive codebooks that minimize the self-interference coupled between each transmit and receive beam pair while ensuring the codebooks remain useful in delivering service to users.
\begin{subequations} \label{eq:problem-desired-i}
\begin{align}
\min_{\mF,\mW} \ & \ \normfro{\mW\ctrans \mH \mF}^2 \\
\st 
& \ \normtwo{\Nt\cdot\vone - \diag{\Atx\ctrans\mF}}^2 \leq \sigmatx^2 \cdot \Nt^2 \cdot \Mtx \label{eq:constraint-cov-tx} \\
& \ \normtwo{\Nr\cdot\vone - \diag{\Arx\ctrans\mW}}^2 \leq \sigmarx^2 \cdot \Nr^2 \cdot \Mrx \label{eq:constraint-cov-rx} \\
& \ \entry{\mF}{:,i} \in \precb \ \forall \ i = 1, \dots, \Mtx \label{eq:constraint-tx-limited} \\
& \ \entry{\mW}{:,j} \in \comcb \ \forall \ j = 1, \dots, \Mrx \label{eq:constraint-rx-limited}
\end{align}
\end{subequations}
By minimizing the Frobenius norm in the objective, our designed codebooks aim to minimize the average self-interference coupled across all transmit-receive beam pairs as evident in \eqref{eq:inr-rx-avg}.
While doing so, our codebooks must satisfy the coverage constraints \eqref{eq:constraint-cov-tx} and \eqref{eq:constraint-cov-rx} that were established in \eqref{eq:coverage-constraint-tx} and \eqref{eq:coverage-constraint-rx}.
Constraints \eqref{eq:constraint-tx-limited} and \eqref{eq:constraint-rx-limited} require our design to create codebooks whose beams are physically realizable with digitally-controlled phase shifters and attenuators. 

Practical systems cannot solve problem \eqref{eq:problem-desired-i} since they do not have knowledge of the true self-interference channel $\mH$ but rather the estimate $\mHbar$.
This motivates us to modify problem \eqref{eq:problem-desired-i} to incorporate robustness to channel estimation error.
It is common in robust communication system optimization to minimize the worst-case error inflicted by some uncertainty \cite{cvx_boyd,convex_beamforming_2010,palomar_robust_2008,palomar_robust_2009}.
In our case, $\mDelta$ has \iid Gaussian entries, meaning the worst-case error inflicted on the objective of problem \eqref{eq:problem-desired-i} is unbounded.
In light of this, we instead minimize the expectation of our objective over the distribution of $\mH$.
Forming problem \eqref{eq:problem-desired-ii}, \lonestar aims to minimize the {expected} average self-interference coupled between beam pairs, given some estimate $\mHbar$. 
\begin{subequations} \label{eq:problem-desired-ii}
\begin{align}
\min_{\mF,\mW} \ & \ \expect_{\mH} \brackets{\normfro{\mW\ctrans \mH \mF}^2 \ \vert \ \mHbar} \label{eq:problem-desired-ii-obj} \\
\st 
& \ \normtwo{\Nt\cdot\vone - \diag{\Atx\ctrans\mF}}^2 \leq \sigmatx^2 \cdot \Nt^2 \cdot \Mtx \\
& \ \normtwo{\Nr\cdot\vone - \diag{\Arx\ctrans\mW}}^2 \leq \sigmarx^2 \cdot \Nr^2 \cdot \Mrx \\
& \ \entry{\mF}{:,i} \in \precb \ \forall \ i = 1, \dots, \Mtx \\
& \ \entry{\mW}{:,j} \in \comcb \ \forall \ j = 1, \dots, \Mrx
\end{align}
\end{subequations}
To transform problem \eqref{eq:problem-desired-ii} into one we can more readily solve, we derive a closed-form expression of the objective \eqref{eq:problem-desired-ii-obj}, courtesy of the fact that $\mDelta$ is Gaussian.

\begin{theorem} \label{thm:objective-expectation}
    The objective of problem \eqref{eq:problem-desired-ii}
    is equivalent to
    \begin{align}
    \expect_{\mH} \brackets{\normfro{\mW\ctrans \mH \mF}^2 \ \vert \ \mHbar} =
    \normfro{\mW\ctrans \mHbar \mF}^2 + \epsilon^2 \cdot \normfro{\mF}^2 \cdot \normfro{\mW}^2.
    \end{align}
    \begin{proof}
        We begin by recalling \eqref{eq:channel} to write
        \begin{align}
        \expect_{\mH} \brackets{\normfro{\mW\ctrans \mH \mF}^2 \ \vert \ \mHbar}
        =& \
        \expect_{\mDelta} \brackets{\normfro{\mW\ctrans \parens{\mHbar + \mDelta} \mF}^2} \\
        \overset{\textrm{(a)}}{=}& \ \expect_{\mDelta} \brackets{\trace{\parens{\mW\ctrans \parens{\mHbar + \mDelta} \mF} \parens{\mW\ctrans \parens{\mHbar + \mDelta} \mF}\ctrans}} \\
        \overset{\textrm{(b)}}{=}& \ \normfro{\mW\ctrans \mHbar \mF}^2 + \trace{\mF\ctrans \expect_{\mDelta} \brackets{\mDelta\ctrans \mW \mW\ctrans \mDelta} \mF} \\
        \overset{\textrm{(c)}}{=}& \ \normfro{\mW\ctrans \mHbar \mF}^2 + \trace{\mF\ctrans \expect_{\mDelta} \brackets{\mDelta\ctrans \mU_{\mW} \mLambda_{\mW} \mU_{\mW}\ctrans \mDelta} \mF} \\
        \overset{\textrm{(d)}}{=}& \ \normfro{\mW\ctrans \mHbar \mF}^2 + \trace{\mF\ctrans \expect_{\tilde\mDelta} \brackets{\tilde\mDelta \mLambda_{\mW} \tilde\mDelta\ctrans} \mF} \\
        \overset{\textrm{(e)}}{=}& \ \normfro{\mW\ctrans \mHbar \mF}^2 + \trace{\mF\ctrans \mF} \cdot \epsilon^2 \cdot \trace{\mLambda_{\mW}} \\
        \overset{\textrm{(f)}}{=}& \ \normfro{\mW\ctrans \mHbar \mF}^2 + \epsilon^2 \cdot \normfro{\mF}^2 \cdot \normfro{\mW}^2
        \end{align}
        where (a) follows from $\normfro{\mA}^2 = \trace{\mA\mA\ctrans}$; (b) is by the linearity of expectation and trace and the fact that $\ev{\mDelta} = \zeromat$; (c) is by taking the  eigenvalue decomposition $\mW\mW\ctrans = \mU_{\mW} \mLambda_{\mW} \mU_{\mW}\ctrans$; (d) is by defining $\tilde\mDelta = \mDelta\ctrans \mU_{\mW}$, which has the same distribution as $\mDelta$ by unitary invariance of \iid zero-mean Gaussian matrices \cite{heath_lozano}; (e) is via the property \cite{gupta_matrix}
        \begin{align}
        \expect_{\mX} \brackets{\mX \mA \mX\ctrans} = \epsilon^2 \cdot \trace{\mA\ctrans} \cdot \mI
        \end{align}
        where $\mX$ has entries \iid Gaussian with mean $0$ and variance $\epsilon^2$; and (f) follows from the definition of Frobenius norm.
    \end{proof}
\end{theorem}

Using \thmref{thm:objective-expectation}, problem \eqref{eq:problem-desired-ii} can be rewritten equivalently as follows. 
\begin{subequations} \label{eq:problem-desired-final}
\begin{align}
\min_{\mF,\mW} \ & \ \normfro{\mW\ctrans \mHbar \mF}^2 + \epsilon^2 \cdot \normfro{\mF}^2 \cdot \normfro{\mW}^2 \\
\st  
& \ \normtwo{\Nt\cdot\vone - \diag{\Atx\ctrans\mF}}^2 \leq \sigmatx^2 \cdot \Nt^2 \cdot \Mtx \\
& \ \normtwo{\Nr\cdot\vone - \diag{\Arx\ctrans\mW}}^2 \leq \sigmarx^2 \cdot \Nr^2 \cdot \Mrx \\
& \ \entry{\mF}{:,i} \in \precb \ \forall \ i = 1, \dots, \Mtx \\
& \ \entry{\mW}{:,j} \in \comcb \ \forall \ j = 1, \dots, \Mrx
\end{align}
\end{subequations}
For some channel estimate $\mHbar$ and knowledge of the estimation error variance $\eps^2$, engineers can aim to solve problem \eqref{eq:problem-desired-final} for codebooks $\mF$ and $\mW$ that minimize the expected average self-interference coupled between transmit and receive beams while ensuring coverage is delivered with some quality parameterized by $\sigmatx^2$ and $\sigmarx^2$.
This concludes the formulation of our codebook design problem.
In the next section, we present a means to solve problem \eqref{eq:problem-desired-final}, along with a summary of and commentary on \lonestar.
\section{Solving for \lonestar Codebooks and Design Remarks} \label{sec:contribution-ii}

In the previous section, we formulated problem \eqref{eq:problem-desired-final}, a codebook design problem that accounts for self-interference channel estimation error and digitally-controlled phase shifters and attenuators.
In this section, we present an approach for (approximately) solving this design problem using a projected alternating minimization approach.
We conclude this section with a summary of our design and remarks regarding design decisions and future research directions.  

Problem \eqref{eq:problem-desired-final} cannot be efficiently solved directly due to the non-convexity posed by digitally-controlled phase shifters and attenuators (i.e., $\precb$ and $\comcb$ are non-convex).
The joint optimization of $\mF$ and $\mW$ introduces further complexity, especially given their dimensionality.
In order to approximately but efficiently solve problem \eqref{eq:problem-desired-final}, we take a projected alternating minimization approach, which is summarized in \algref{alg:algorithm-ii}.
Let $\project{\entry{\mF}{:,i}}{\precb}$ be the projection of the $i$-th beam (column) of $\mF$ onto the set of physically realizable beamforming vectors $\precb$.
We extend this projection onto each beam in $\mF$ as $\project{\mF}{\precb}$ by applying it column-wise on $\mF$; the projection of $\mW$ onto $\comcb$ is defined analogously.
With digitally-controlled phase shifters and attenuators, each element of $\mF$ must be quantized in both phase and amplitude.
Since the set of physically realizable beamforming weights is the same for each element in a beamforming vector, the projection of that vector reduces to projecting each of its elements.
The $(m,n)$-th element following the projection of $\mF$ onto $\precb$ can be expressed as
\begin{align} \label{eq:project}
\entry{\project{\mF}{\precb}}{m,n} = A\opt \cdot \exp\parens{\j \cdot \theta\opt}
\end{align}
where $A\opt$ and $\theta\opt$ are the quantized amplitude and phase of $\entry{\mF}{m,n}$.
These can be expressed as
\begin{align}
A\opt &= \argmin_{A_i \in \set{A}} \ \bars{A_i - A}, \qquad
\theta\opt = \argmin_{\theta_i \in \set{P}} \ \bars{\exp\parens{\j \cdot \theta_i} - \exp\parens{\j \cdot \theta}}
\end{align}
where $A = \bars{\entry{\mF}{m,n}}$ and $\theta = \angleop{\entry{\mF}{m,n}}$ are the magnitude and phase of $\entry{\mF}{m,n}$.
Here, $\set{A}$ and $\set{P}$ are the sets of physically realizable amplitudes (attenuator settings) and phases (phase shifter settings), respectively.

%
%

To approximately solve problem \eqref{eq:problem-desired-final}, we begin with $\precbmat \leftarrow \project{\Atx}{\precb}$ and $\comcbmat \leftarrow \project{\Arx}{\comcb}$, 
which initializes our beams as matched filters to achieve (near) maximum gains across $\dirtxset$ and $\dirrxset$, respectively.
Then, we fix the receive codebook $\comcbmat$ and solve problem \eqref{eq:problem-am-F} below.
\begin{subequations} \label{eq:problem-am-F}
    \begin{align}
    \min_{\mF} \ & \ \normfro{\mW\ctrans \mHbar \mF}^2 + \epsilon^2 \cdot \normfro{\mF}^2 \cdot \normfro{\mW}^2 \\
    \st 
    & \ \normtwo{\Nt\cdot\vone - \diag{\Atx\ctrans\mF}}^2 \leq \sigmatx^2 \cdot \Nt^2 \cdot \Mtx \\
    & \ \bars{\entry{\mF}{i,j}} \leq 1, \ i = 1, \dots, \Nt, \ j = 1, \dots, \Mtx \label{eq:problem-unit-entry-F}
    \end{align}
\end{subequations}
As a temporary convex relaxation, we have introduced constraint \eqref{eq:problem-unit-entry-F} instead of requiring strict quantized phase and amplitude control.
This enforces that each beamforming weight not exceed unit magnitude, understood by the fact that attenuators are used for amplitude control as outlined in \eqref{eq:unit-f-w}.
Problem \eqref{eq:problem-am-F} is convex and can be readily solved using convex optimization solvers (e.g., we used \cite{cvx}).
Dedicated algorithms more efficient than general solvers can likely be developed for \lonestar, but that is beyond the scope of this work, making it a good topic for future work.
After solving problem \eqref{eq:problem-am-F}, the solution $\mF\opt$ is projected onto the set $\precb$ as $\precbmat \leftarrow \project{\mF\opt}{\precb}$ 
to ensure it is physically realizable.
The projected $\precbmat$ is then used when solving for the receive codebook in problem \eqref{eq:problem-am-W} below.
\begin{subequations} \label{eq:problem-am-W}
\begin{align}
\min_{\mW} \ & \ \normfro{\mW\ctrans \mHbar \mF}^2 + \epsilon^2 \cdot \normfro{\mF}^2 \cdot \normfro{\mW}^2 \\
\st 
& \ \normtwo{\Nr\cdot\vone - \diag{\Arx\ctrans\mW}}^2 \leq \sigmarx^2 \cdot \Nr^2 \cdot \Mrx \\
& \ \bars{\entry{\mW}{i,j}} \leq 1, \ i = 1, \dots, \Nr, \ j = 1, \dots, \Mrx
\end{align}
\end{subequations}
Problem \eqref{eq:problem-am-W} has a form identical to problem \eqref{eq:problem-am-F} and is thus convex and can be solved using the same approach used for problem \eqref{eq:problem-am-F}.
Upon solving problem \eqref{eq:problem-am-W}, we ensure the solution $\mW\opt$ is physically realizable by projecting it onto $\comcb$ via $\comcbmat \leftarrow \project{\mW\opt}{\comcb}$. 
This concludes our design of \lonestar, which has produced physically realizable transmit and receive codebook matrices $\precbmat$ and $\comcbmat$.
These can then be unpacked via Definition~\ref{def:codebooks} to form the sets $\precbbar$ and $\comcbbar$ that can be used for beam alignment and subsequent full-duplexing of downlink and uplink. 
In the next section, we evaluate \lonestar extensively through simulation.

\begin{algorithm}[!t]
    \begin{algorithmic}[0]
        \STATE {1. Initialize $\precbmat \leftarrow \project{\Atx}{\precb}$ and
            $\comcbmat \leftarrow \project{\Arx}{\comcb}$.}
        \STATE {2. Solve problem \eqref{eq:problem-am-F} for the transmit codebook $\mF\opt$.}
         \STATE {3. Project $\mF\opt$ to ensure it is physically realizable:~$\mF \leftarrow \project{\mF\opt}{\precb}$.}
        \STATE {4. Solve problem \eqref{eq:problem-am-W} for the receive codebook $\mW\opt$ using the updated $\mF$.}
        \STATE {5. Project $\mW\opt$ to ensure it is physically realizable:~$\mW \leftarrow \project{\mW\opt}{\comcb}$.}
    \end{algorithmic}
    \caption{Projected alternating minimization to approximately solve problem \eqref{eq:problem-desired-final}.}
    \label{alg:algorithm-ii}
\end{algorithm}




\comment{
\begin{remark}[Summary]
A summary of \lonestar is outlined in \algref{alg:algorithm-i}.
\lonestar begins by obtaining an estimate $\mHbar$ of the self-interference \mimo channel and knowledge of the channel estimation error variance $\eps^2$.
Then, the transmit and receive coverage regions $\txdirsetcb$ and $\rxdirsetcb$ are defined by system engineers, and the array response matrices $\Atx$ and $\Arx$ are assembled based on the transmit and receive array geometries.
Engineers specify the tolerated transmit and receive coverage variances $\sigmatx^2$ and $\sigmarx^2$, and \lonestar codebook matrices $\mF$ and $\mW$ are designed by solving problem \eqref{eq:problem-desired-final} using \algref{alg:algorithm-ii}.
Codebooks $\precbbar$ and $\comcbbar$ are formed by unpacking $\mF$ and $\mW$.
The \bs then uses $\precbbar$ and $\comcbbar$ for beam alignment and full-duplexing of downlink and uplink over several time slots before rerunning \lonestar to account for drift in the self-interference channel.
\end{remark}
}

\begin{remark}[Design decisions]
    We present a summary of \lonestar in \algref{alg:algorithm-i}, and provide a few remarks on our approach and communicate our motivations and design decisions.
    The objective of \lonestar is to create transmit and receive codebooks that minimize the average $\inrrx$ while still delivering high transmit and receive beamforming gain.
    It may seem preferred to maximize the average $\sinrrx$ or the sum spectral efficiency, but this requires knowledge of the transmit and receive \gpsnr, which the system only has \textit{after} conducting beam alignment, or real-time knowledge of $\vhtx$ and $\vhrx$, which is currently impractical.
    \lonestar only relies on knowledge of the self-interference \mimo channel and not that of the transmit and receive channels.
    By minimizing $\inrrx$ while constraining transmit and receive coverage quality (i.e., maintaining high $\snrtx$ and $\snrrx$), \lonestar can consequently achieve high $\sinrrx$ and $\sinrtx$.
    One may also suggest we instead minimize the codebooks' maximum $\inrrx$ (to minimize worst-case self-interference) rather than minimize the average $\inrrx$.
    We chose to minimize the average $\inrrx$ because this prevents \lonestar from being significantly hindered if there are select transmit-receive beam pairs that simply cannot offer low self-interference while delivering high beamforming gain.
    It is important to note that the Frobenius norm objective used by \lonestar will make it more incentivized to reduce high self-interference but does not exclusively prioritize these worst-case beam pairs  \cite{cvx_boyd}.
    Individual beam pairs that do not yield sufficiently low $\inrrx$ when minimizing the average $\inrrx$ with \lonestar could potentially be avoided via scheduling or half-duplexed as a last resort.
\end{remark}

\begin{remark}[Complexity]
    It is difficult to comment on the complexity of our design for a few reasons.
    First and foremost, its complexity depends heavily on the solver used for solving problems \eqref{eq:problem-am-F} and \eqref{eq:problem-am-W}.
    As mentioned, more efficient solving algorithms may be tailored for \lonestar, but that is out of the scope of this work.
    Additionally, total computational costs will depend on the initialization of $\mF$ and $\mW$.
    We have initialized them both as conjugate beamforming codebooks; in practice, it would likely be efficient to use previous solutions of $\mF$ and $\mW$ when rerunning \lonestar to account for drift in the self-interference channel.
    Note that our design takes place at the full-duplex \bs, which presumably has sufficient compute resources to execute \lonestar.
    Practically, it may be preferred to conduct a thorough estimation of $\mH$ and execution of \lonestar during initial setup of the \bs and then perform updates to the estimate of $\mH$ and possibly to our design over time; this would be interesting future work.
    Most existing beamforming-based full-duplex solutions need to be executed for each downlink-uplink user pair, whereas the same \lonestar design can be used across many user pairs, meaning it may offer significant computational savings in comparison. 
\end{remark}

\begin{algorithm}[!t]
    \begin{algorithmic}[0]
        \STATE {1. Obtain an estimate $\mHbar$ of the self-interference channel and the estimation error variance $\eps^2$.}
        \STATE {2. Define transmit and receive coverage regions $\txdirsetcb$ and $\rxdirsetcb$ and assemble $\Atx$ and $\Arx$.}
        \STATE {3. Set transmit and receive coverage variances $\sigmatx^2$ and $\sigmarx^2$.}
        \STATE {4. Solve problem \eqref{eq:problem-desired-final} for $\precbmat$ and $\comcbmat$ using \algref{alg:algorithm-ii}.}
        \STATE {5. Unpack $\precbmat$ and $\comcbmat$ to form codebooks $\precbbar$ and $\comcbbar$ using Definition~\ref{def:codebooks}.} 
    \end{algorithmic}
    \caption{A summary of our analog beamforming codebook design, \lonestar.}
    \label{alg:algorithm-i}
\end{algorithm}

\comment{
\begin{remark}[Future directions]
We now outline future research directions particularly relevant to \lonestar.
Extensive study and modeling of \mmwave self-interference channels and their coherence time will be important to understanding the potentials of \lonestar---and full-duplex \mmwave systems in general---along with developing efficient means for self-interference channel estimation.
Also, methods to dynamically update \lonestar as the self-interference channel changes would be valuable future work, along with implementations of \lonestar using real \mmwave platforms.
Naturally, there are some ways to improve upon this work, such as by better handling limited phase and amplitude control and by creating efficient, dedicated solvers for \lonestar.
Beyond \lonestar, measurement-driven beamforming-based full-duplex solutions (potentially using machine learning) that do not require self-interference \mimo channel knowledge (e.g., \steer \cite{roberts_steer}) would be excellent contributions to the research community.
\end{remark}
}

\comment{
---

\begin{align}
\min_{\mF,\mW} \ & \ \maxop{\frac{\powertxbs \cdot G^2}{\powernoisebs} \cdot \frac{\normfro{\mW\ctrans \mH \mF}^2}{\Mtx \cdot \Mrx},\inrrxthresh} \\
\st \ 
& \ \normtwo{\Nt\cdot\vone - \diag{\Atx\ctrans\mF}}^2 \leq \sigmatx^2 \cdot \Nt^2 \cdot \Mtx \\
& \ \normtwo{\Nr\cdot\vone - \diag{\Arx\ctrans\mW}}^2 \leq \sigmarx^2 \cdot \Nr^2 \cdot \Mrx \\
& \ \entry{\mF}{:,i} \in \precb \ \forall \ i = 1, \dots, \Mtx \\
& \ \entry{\mW}{:,j} \in \comcb \ \forall \ j = 1, \dots, \Mrx
\end{align}

\begin{align}
\gamma = \sqrt{\frac{\inrrxthresh \cdot \Mtx \cdot \Mrx \cdot \powernoisebs}{\powertxbs \cdot G^2}}
\end{align}

\begin{subequations} \label{eq:problem-iii}
    \begin{align}
    \min_{\mF,\mW} \ & \ \maxop{\normfro{\mW\ctrans \mH \mF},\gamma} \\
    \st \ 
    & \ \normtwo{\Nt\cdot\vone - \diag{\Atx\ctrans\mF}}^2 \leq \sigmatx^2 \cdot \Nt^2 \cdot \Mtx \\
    & \ \normtwo{\Nr\cdot\vone - \diag{\Arx\ctrans\mW}}^2 \leq \sigmarx^2 \cdot \Nr^2 \cdot \Mrx \\
    & \ \entry{\mF}{:,i} \in \precb \ \forall \ i = 1, \dots, \Mtx \label{eq:problem-quantize-tx-1} \\
    & \ \entry{\mW}{:,j} \in \comcb \ \forall \ j = 1, \dots, \Mrx \label{eq:problem-quantize-rx-1}
    \end{align}
\end{subequations}

\begin{subequations} \label{eq:problem-iii}
    \begin{align}
    \min_{\mF,\mW} \ & \ \maxop{\normfro{\mW\ctrans \parens{\mHbar + \mDelta} \mF},\gamma} \\
    \st \ 
    & \ \normtwo{\Nt\cdot\vone - \diag{\Atx\ctrans\mF}}^2 \leq \sigmatx^2 \cdot \Nt^2 \cdot \Mtx \\
    & \ \normtwo{\Nr\cdot\vone - \diag{\Arx\ctrans\mW}}^2 \leq \sigmarx^2 \cdot \Nr^2 \cdot \Mrx \\
    & \ \entry{\mF}{:,i} \in \precb \ \forall \ i = 1, \dots, \Mtx \label{eq:problem-quantize-tx-1} \\
    & \ \entry{\mW}{:,j} \in \comcb \ \forall \ j = 1, \dots, \Mrx \label{eq:problem-quantize-rx-1}
    \end{align}
\end{subequations}

\begin{subequations} \label{eq:problem-iii}
    \begin{align}
    \min_{\mF,\mW} \ & \ \maxop{\max_{\mDelta} \ \normfro{\mW\ctrans \parens{\mHbar + \mDelta} \mF},\gamma} \\
    \st \ 
    & \ \normtwo{\Nt\cdot\vone - \diag{\Atx\ctrans\mF}}^2 \leq \sigmatx^2 \cdot \Nt^2 \cdot \Mtx \\
    & \ \normtwo{\Nr\cdot\vone - \diag{\Arx\ctrans\mW}}^2 \leq \sigmarx^2 \cdot \Nr^2 \cdot \Mrx \\
    & \ \entry{\mF}{:,i} \in \precb \ \forall \ i = 1, \dots, \Mtx \label{eq:problem-quantize-tx-1} \\
    & \ \entry{\mW}{:,j} \in \comcb \ \forall \ j = 1, \dots, \Mrx \label{eq:problem-quantize-rx-1} \\
    & \ \normfro{\mDelta}^2 \leq \eps^2 \cdot \Nt \cdot \Nr
    \end{align}
\end{subequations}

\begin{subequations} \label{eq:problem-iii}
    \begin{align}
    \min_{\mF,\mW} \ & \ \maxop{\normfro{\mW\ctrans \mHbar \mF} + \eps \cdot \svmax{\mW} \cdot \svmax{\mF},\gamma} \\
    \st \ 
    & \ \normtwo{\Nt\cdot\vone - \diag{\Atx\ctrans\mF}}^2 \leq \sigmatx^2 \cdot \Nt^2 \cdot \Mtx \\
    & \ \normtwo{\Nr\cdot\vone - \diag{\Arx\ctrans\mW}}^2 \leq \sigmarx^2 \cdot \Nr^2 \cdot \Mrx \\
    & \ \entry{\mF}{:,i} \in \precb \ \forall \ i = 1, \dots, \Mtx \label{eq:problem-quantize-tx-1} \\
    & \ \entry{\mW}{:,j} \in \comcb \ \forall \ j = 1, \dots, \Mrx \label{eq:problem-quantize-rx-1}
    \end{align}
\end{subequations}

---

\begin{align}
\Nt^2 \cdot \Nr^2 = \max_{\vf,\vw,\mH} \ & \ \bars{\vw\ctrans \mH \vf}^2 \\
\st \ 
& \ \normfro{\mH}^2 = \Nt \cdot \Nr \\
& \ \normtwo{\vf}^2 = \Nt \\
& \ \normtwo{\vw}^2 = \Nr
\end{align}

\begin{align}
\svmaxsq{\mH} \cdot \Nt \cdot \Nr = \max_{\vf,\vw} \ & \ \bars{\vw\ctrans \mH \vf}^2 \\
\st \ 
& \ \normtwo{\vf}^2 = \Nt \\
& \ \normtwo{\vw}^2 = \Nr
\end{align}

So max \ginr is 
\begin{align}
\inrrxbar = \frac{\powertxbs \cdot G^2 \cdot \Nt^2 \cdot \Nr^2}{\powernoisebs} \geq \inrrx
\end{align}

INR target $\inrrxthresh$
\begin{align}
\inrrxthresh = \frac{\powertxbs \cdot G^2 \cdot \gamma}{\powernoisebs}
\end{align}

The average \ginr coupled across all beam pairs is
\begin{align}
\frac{\powertxbs \cdot G^2}{\powernoisebs} \cdot \frac{\normfro{\mW\ctrans \mH \mF}^2}{\Mtx \cdot \Mrx}
\end{align}

We aim to minimize the average \ginr coupled across all beam pairs but do not incentivize it to reduce such below some target $\inrrxthresh$.
\begin{subequations} \label{eq:problem-i}
    \begin{align}
    \min_{\mF,\mW} \ & \ \maxop{\frac{\powertxbs \cdot G^2}{\powernoisebs} \cdot \frac{\normfro{\mW\ctrans \mH \mF}^2}{\Mtx \cdot \Mrx},\inrrxthresh} \\
    \st \ 
    & \ \normtwo{\Nt\cdot\vone - \diag{\Atx\ctrans\mF}}^2 \leq \sigmatx^2 \cdot \Nt^2 \cdot \Mtx \\
    & \ \normtwo{\Nr\cdot\vone - \diag{\Arx\ctrans\mW}}^2 \leq \sigmarx^2 \cdot \Nr^2 \cdot \Mrx \\
    & \ \entry{\mF}{:,i} \in \precb \ \forall \ i = 1, \dots, \Mtx \label{eq:problem-quantize-tx-1} \\
    & \ \entry{\mW}{:,j} \in \comcb \ \forall \ j = 1, \dots, \Mrx \label{eq:problem-quantize-rx-1}
    \end{align}
\end{subequations}

\begin{subequations} \label{eq:problem-ii}
\begin{align}
\min_{\mF,\mW} \ & \ \maxop{\frac{\powertxbs \cdot G^2}{\powernoisebs} \cdot \frac{\normfro{\mW\ctrans \parens{\mHbar + \mDelta} \mF}^2}{\Mtx \cdot \Mrx},\inrrxthresh} \\
\st \ 
& \ \normtwo{\Nt\cdot\vone - \diag{\Atx\ctrans\mF}}^2 \leq \sigmatx^2 \cdot \Nt^2 \cdot \Mtx \\
& \ \normtwo{\Nr\cdot\vone - \diag{\Arx\ctrans\mW}}^2 \leq \sigmarx^2 \cdot \Nr^2 \cdot \Mrx \\
& \ \entry{\mF}{:,i} \in \precb \ \forall \ i = 1, \dots, \Mtx \label{eq:problem-quantize-tx-1} \\
& \ \entry{\mW}{:,j} \in \comcb \ \forall \ j = 1, \dots, \Mrx \label{eq:problem-quantize-rx-1}
\end{align}
\end{subequations}

Minimize the worst-case average \ginr caused by channel estimation error.
\begin{subequations} \label{eq:problem-ii}
    \begin{align}
    \min_{\mF,\mW} \ & \ \maxop{\max_{\mDelta} \frac{\powertxbs \cdot G^2}{\powernoisebs} \cdot \frac{\normfro{\mW\ctrans \parens{\mHbar + \mDelta} \mF}^2}{\Mtx \cdot \Mrx},\inrrxthresh} \\
    \st \ 
    & \ \normtwo{\Nt\cdot\vone - \diag{\Atx\ctrans\mF}}^2 \leq \sigmatx^2 \cdot \Nt^2 \cdot \Mtx \\
    & \ \normtwo{\Nr\cdot\vone - \diag{\Arx\ctrans\mW}}^2 \leq \sigmarx^2 \cdot \Nr^2 \cdot \Mrx \\
    & \ \entry{\mF}{:,i} \in \precb \ \forall \ i = 1, \dots, \Mtx \label{eq:problem-quantize-tx-1} \\
    & \ \entry{\mW}{:,j} \in \comcb \ \forall \ j = 1, \dots, \Mrx \label{eq:problem-quantize-rx-1} \\
    & \ \normfro{\mDelta}^2 \leq \eps^2 \cdot \Nt \cdot \Nr
    \end{align}
\end{subequations}

\begin{align}
\max_{\normfro{\mDelta}^2 \leq \eps^2} \ \normfro{\mW\ctrans \parens{\mHbar + \sqrt{\Nt \cdot \Nr} \cdot \mDelta} \mF}
&= \max_{\normfro{\mDelta}^2 \leq \eps^2} \ \normfro{\mW\ctrans \mHbar \mF + \sqrt{\Nt \cdot \Nr} \cdot \mW\ctrans \mDelta \mF} \\
&= \normfro{\mW\ctrans \mHbar \mF} + \sqrt{\Nt \cdot \Nr} \cdot \eps \cdot \svmax{\mW} \cdot \svmax{\mF}
\end{align}

\begin{subequations} \label{eq:problem-iii}
    \begin{align}
    \min_{\mF,\mW} \ & \ \maxop{\normfro{\mW\ctrans \mHbar \mF} + \eps \cdot \sqrt{\Nt \cdot \Nr} \cdot \svmax{\mW} \cdot \svmax{\mF},\gamma} \\
    \st \ 
    & \ \normtwo{\Nt\cdot\vone - \diag{\Atx\ctrans\mF}}^2 \leq \sigmatx^2 \cdot \Nt^2 \cdot \Mtx \\
    & \ \normtwo{\Nr\cdot\vone - \diag{\Arx\ctrans\mW}}^2 \leq \sigmarx^2 \cdot \Nr^2 \cdot \Mrx \\
    & \ \entry{\mF}{:,i} \in \precb \ \forall \ i = 1, \dots, \Mtx \label{eq:problem-quantize-tx-1} \\
    & \ \entry{\mW}{:,j} \in \comcb \ \forall \ j = 1, \dots, \Mrx \label{eq:problem-quantize-rx-1}
    \end{align}
\end{subequations}

\section{Extra}

\begin{align}
\min_{\mF,\mW} \ & \ \maxop{\normfro{\mW\ctrans \mH \mF}^2,\gamma} \\
\st \ 
& \ \normtwo{\Nt\cdot\vone - \diag{\Atx\ctrans\mF}}^2 \leq \sigmatx^2 \cdot \Nt^2 \cdot \Mtx \\
& \ \normtwo{\Nr\cdot\vone - \diag{\Arx\ctrans\mW}}^2 \leq \sigmarx^2 \cdot \Nr^2 \cdot \Mrx \\
& \ \entry{\mF}{:,i} \in \precb \ \forall \ i = 1, \dots, \Mtx \\
& \ \entry{\mW}{:,j} \in \comcb \ \forall \ j = 1, \dots, \Mrx
\end{align}
}

\section{Simulation Setup and Performance Metrics} \label{sec:simulation-setup}


We simulated the full-duplex system illustrated in \figref{fig:system} at 30 GHz in a Monte Carlo fashion to extensively evaluate \lonestar against conventional codebooks \cite{mfm_arxiv}.
The full-duplex \bs is equipped with two $8 \times 8$ half-wavelength \acrlongpl{upa} for transmission and reception; users have a single antenna for simplicity.
Analog beamforming networks at the \bs use log-stepped attenuators with $0.5$ dB of attenuation per \gls{lsb}.
To serve downlink and uplink users, the full-duplex \bs uses $\Mtx = \Mrx = 45$ beams distributed from $-60^\circ$ to $60^\circ$ in azimuth and from $-30^\circ$ to $30^\circ$ in elevation, each in $15^\circ$ steps.
Users are distributed uniformly from $-67.5^\circ$ to $67.5^\circ$ in azimuth and from $-37.5^\circ$ to $37.5^\circ$ in elevation.
The downlink and uplink channels are simulated as \gls{los} channels for simplicity; similar conclusions would be expected when modeling them otherwise due to the directional nature of \mmwave channels.
To select which transmit and receive beams serve a given downlink and uplink user pair, we use exhaustive beam sweeping on each link independently and choose the beams that maximize $\snrtx$ and $\snrrx$.
The choice of a particular self-interference channel model is a difficult one to justify practically, given there lacks a well-accepted, measurement-backed model.
We have chosen to simulate the self-interference channel $\mH$ using the spherical-wave channel model \cite{spherical_2005}, which captures idealized near-field propagation between the transmit and receive arrays of the full-duplex device and has been used widely thus far in related literature.
This channel model is a function of the relative geometry between the transmit and receive arrays and is described as
\begin{align}
\entry{\mH}{m,n} = \frac{\rho}{r_{n,m}}\exp \parens{-\j 2 \pi \frac{r_{n,m}}{\lambda}} \label{eq:spherical-wave}
\end{align}
where $r_{n,m}$ is the distance between the $n$-th transmit antenna and the $m$-th receive antenna, $\lambda$ is the carrier wavelength, and $\rho$ is a normalizing factor to satisfy $\normfro{\mH}^2 = \Nt \cdot \Nr$.
To realize such a channel, we have vertically stacked our transmit and receive arrays with $10 \lambda$ of separation.
While this channel model may not hold perfectly in practice, it provides a sensible starting point to evaluate \lonestar, particularly when self-interference is dominated by idealized near-field propagation. 
Shortly, for more extensive evaluation, we consider a self-interference channel model that mixes this spherical-wave model with a Rayleigh faded channel.
We do not model cross-link interference explicitly but instead evaluate it over various $\inrtx$.



\subsection{Baselines and Performance Metrics}
Considering \lonestar is the first known codebook design specifically for \mmwave full-duplex, we evaluate it against two conventional codebooks, both of which are used in practice by half-duplex systems to form sets of beams that provide broad coverage with high gain:
\begin{itemize}
    \item A \textit{CBF codebook} that uses conjugate beamforming \cite{heath_lozano}, where
    \begin{align}
    \vf_i = \project{\atx{\dirtx\idx{i}}}{\precb}, \qquad \vw_j = \project{\arx{\dirrx\idx{j}}}{\comcb}.
    \end{align}
    \cbf codebooks can deliver (approximately) maximum gain to all directions in $\txdirsetcb$ and $\rxdirsetcb$.
    The projections here are simply to ensure the beams are physically realizable and do not have significant impacts on the beam shapes with modest resolutions $\bitsphase$ and $\bitsamp$.
    \item A \textit{Taylor codebook} that uses conjugate beamforming with side lobe suppression as
    \begin{align}
    \vf_i = \project{\atx{\dirtx\idx{i}} \odot \vv_{\labeltx}}{\precb}, \qquad \vw_j = \project{\arx{\dirrx\idx{j}} \odot \vv_{\labelrx}}{\comcb}
    \label{eq:cbf-tay}
    \end{align}
    where $\vv_{\labeltx}$ and $\vv_{\labelrx}$ are Taylor windows applied element-wise, providing $25$ dB of side lobe suppression; we found this level of side lobe suppression competes best with \lonestar.
    Taylor windowing is an established means to reduce interference inflicted by side lobe levels at the cost of lessened beamforming gain and a wider main lobe \cite{taylor_1955,taylor_SAR_1995}.
\end{itemize}

We assess \lonestar against our baseline codebooks in terms of sum spectral efficiency of the transmit and receive links.
Specifically, we use the normalized sum spectral efficiency
\begin{align}
\sesumnorm = \frac{\setx + \serx}{\captxcb  + \caprxcb}
\end{align}
where $\captxcb = \logtwo{1 + \snrtxcbf}$ and $\caprxcb = \logtwo{1 + \snrrxcbf}$ we term the \textit{codebook capacities}; 
here, $\snrtxcbf$ and $\snrrxcbf$ are the maximum \gpsnr delivered to a given downlink-uplink user pair by a CBF codebook.
In other words, $\captxcb + \caprxcb$ is the maximum sum spectral efficiency when using a \cbf codebook and interference is inherently not present (i.e., when $\inrrxbar = \inrtx = -\infty$ dB).
With this normalization, $\sesumnorm$ will gauge how well a given codebook delivers transmission and reception in the face of self-interference and cross-link interference, relative to the interference-free capacity when beamforming with a CBF codebook.
Note that $\sesumnorm = 0.5$ can be achieved with a \cbf codebook by half-duplexing transmission and reception with equal \gls{tdd} under an instantaneous power constraint.
As such, \lonestar generally aims to exceed $\sesumnorm = 0.5$ and outperform that achieved by CBF and Taylor codebooks.
Note that $\sesumnorm$ is not truly upper-bounded by $1$ since $\captxcb + \caprxcb$ is not the true sum Shannon capacity; nonetheless, achieving $\sesumnorm \approx 1$ indicates near best-case full-duplex performance one can expect when using beamforming codebooks.
Having simulated our system in a Monte Carlo fashion, in our results we will present $\sesumnorm$ averaged over downlink-uplink user realizations. 

\begin{figure*}
    \centering
    \subfloat[Conventional broadside beams.]{\includegraphics[width=0.475\linewidth,height=\textheight,keepaspectratio]{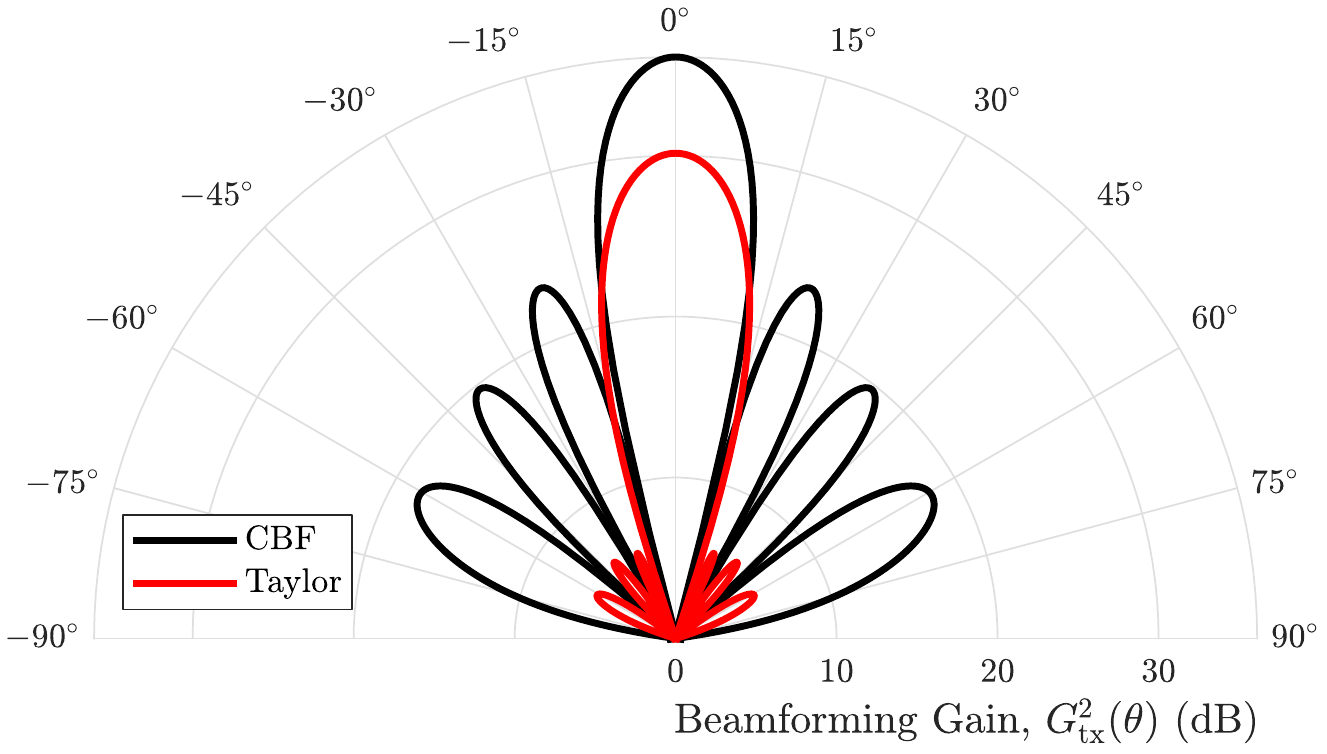}
        \label{fig:beams-a}}
    \quad
    \subfloat[\lonestar beams.]{\includegraphics[width=0.475\linewidth,height=\textheight,keepaspectratio]{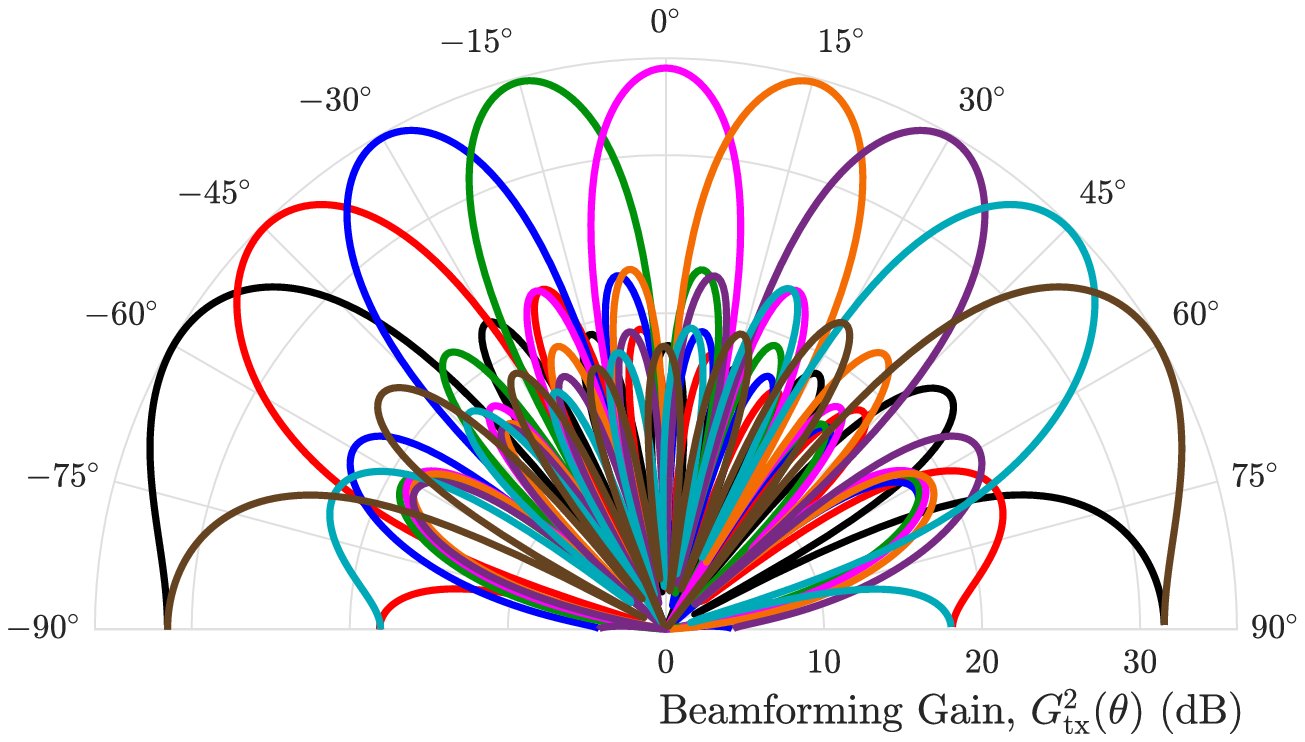}
        \label{fig:beams-b}}
    \caption{(a) The azimuth cut of a broadside beam from conventional \cbf and \taylor codebooks. (b) The azimuth cuts of transmit beams produced by \lonestar that serve various azimuth directions at an elevation of $0^\circ$.} 
    \label{fig:beams}
\end{figure*}

\subsection{Beam Patterns}
In \figref{fig:beams}, we compare conventional beams to those produced by \lonestar.
\figref{fig:beams-a} depicts the shape of beams from our two baseline codebooks.
CBF beams can each deliver maximum gain in their steered direction and exhibit a narrow main lobe and high side lobes.
With the \taylor codebook, these side lobes are reduced to $25$ dB below the main lobe, which itself sacrificed about $6$ dB in maximum gain for this side lobe suppression.
In \figref{fig:beams-b}, we plot transmit beams from a codebook produced by \lonestar; receive beams are similar.
Clearly, unlike the Taylor codebook, \lonestar does not attempt to merely shrink side lobes to reduce self-interference.
Instead,\textbf{ \lonestar makes use of side lobes}, shaping them to cancel self-interference spatially, and while doing so, maintains near maximum gain across its beams.


%
%
%
%
%
%
%

\subsection{Choosing Coverage Variances, $\sigmatx^2$ and $\sigmarx^2$}

\begin{figure*}
    \centering
    \subfloat[As a function of coverage variance, $\sigmatx^2 = \sigmarx^2$.]{\includegraphics[width=\linewidth,height=0.26\textheight,keepaspectratio]{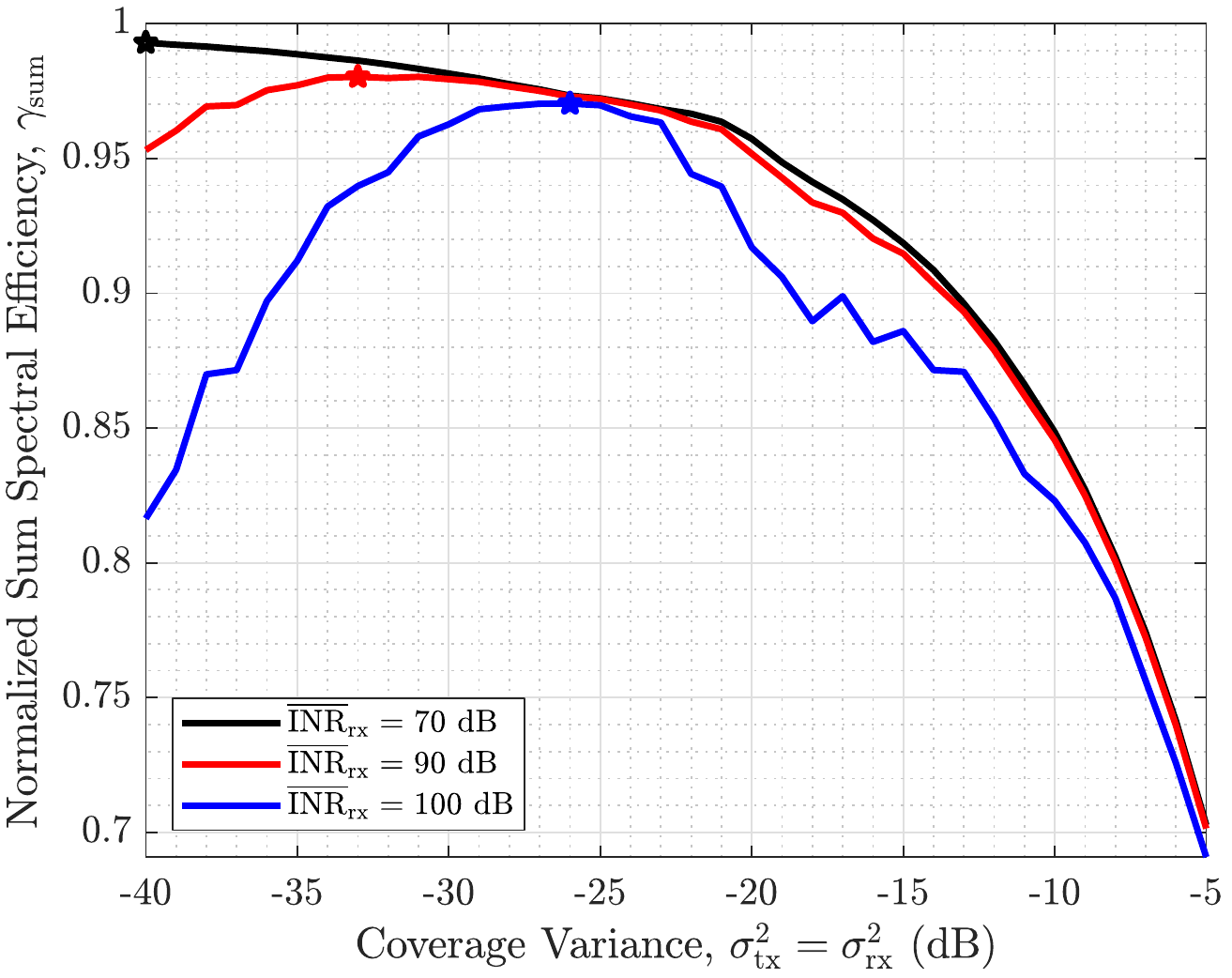}
        \label{fig:choosing-sigma-a}}
    \quad
    \subfloat[As a function of self-interference strength, $\inrrxbar$.]{\includegraphics[width=\linewidth,height=0.26\textheight,keepaspectratio]{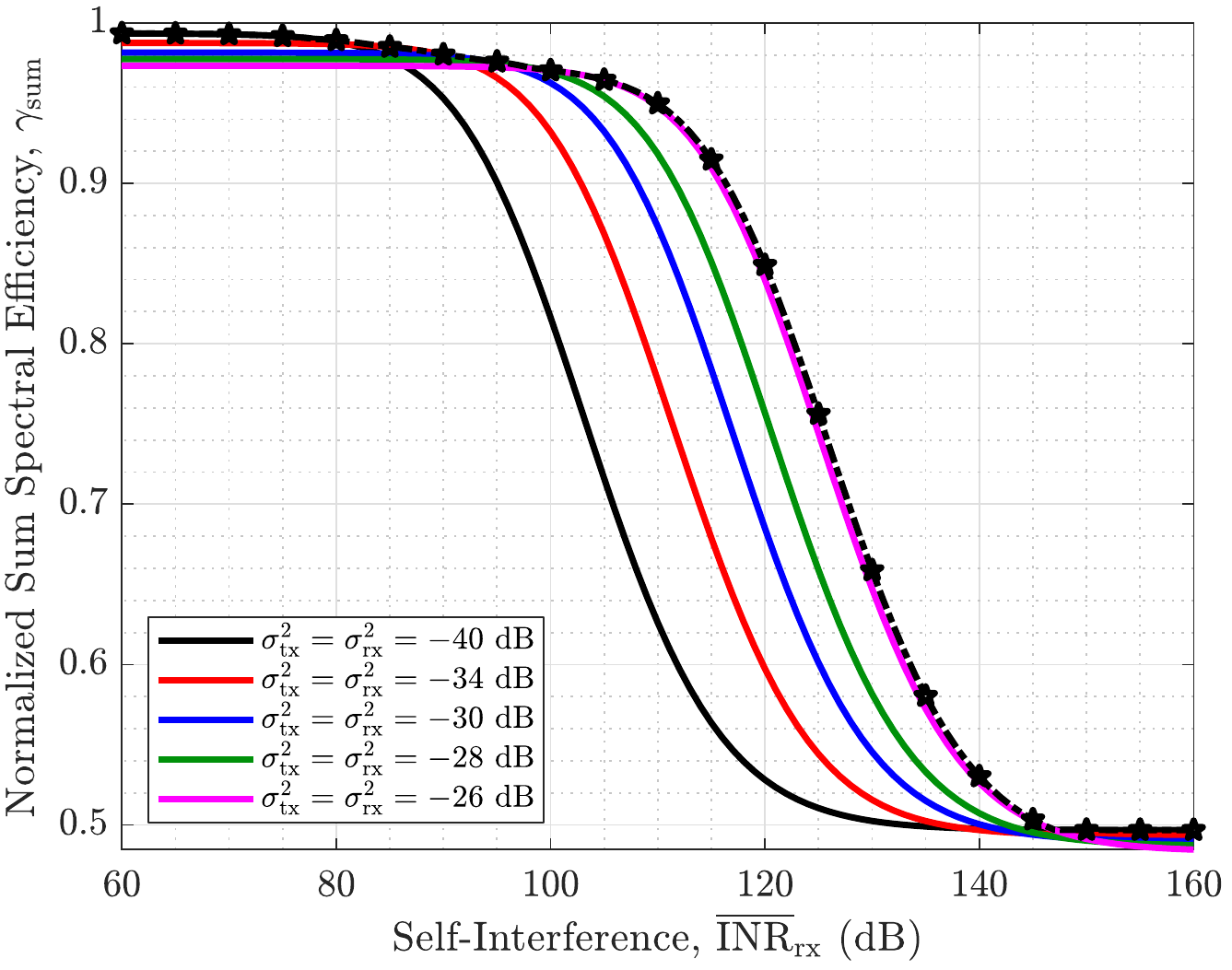}
        \label{fig:choosing-sigma-b}}
    \caption{Normalized sum spectral efficiency $\sesumnorm$ (a) as a function of the coverage variance $\sigmatx^2 = \sigmarx^2$ for various $\inrrxbar$; and (b) as a function of self-interference strength $\inrrxbar$ for various $\sigmatxrxsq$. In both, $\snrtxbar = \snrrxbar = 10$ dB, $\bitsphase = \bitsamp = 8$ bits, $\inrtx = -\infty$ dB, and $\epsilon^2 = -\infty$ dB. Performance with the optimal $\sigmatxrxsq$ is denoted with $\star$.}
    \label{fig:choosing-sigma}
\end{figure*}

In general, it is not possible to analytically state optimal choices for the \lonestar design parameters $\sigmatx^2$ and $\sigmarx^2$ that maximize normalized sum spectral efficiency $\sesumnorm$.
This motivates us to examine heuristically how to choose $\sigmatx^2$ and $\sigmarx^2$.
In \figref{fig:choosing-sigma-a}, we plot the normalized sum spectral efficiency $\sesumnorm$ as a function of $\sigmatx^2 = \sigmarx^2$ for various levels of self-interference $\inrrxbar$.
Here, we let $\snrtxbar = \snrrxbar = 10$ dB and have assumed no channel estimation error nor cross-link interference (i.e., $\eps^2 = -\infty$ dB, $\inrtx = -\infty$ dB).
At low self-interference $\inrrxbar$ (in black), there is inherently low coupling between the transmit and receive beams, meaning it is best to maintain high beamforming gain rather than reduce self-interference further.
This makes a low coverage variance optimal on average, shown as the $\star$ at $\sigmatx^2 = \sigmarx^2 = -40$ dB.
As $\inrrxbar$ increases, \lonestar benefits from having greater transmit and receive coverage variance since it affords more flexibility to reduce self-interference and increase the sum spectral efficiency.

In \figref{fig:choosing-sigma-b}, we highlight this further by plotting the normalized sum spectral efficiency as a function of $\inrrxbar$ for various $\sigmatx^2 = \sigmarx^2$.
Naturally, one may independently choose $\sigmatx^2$ and $\sigmarx^2$, rather than assume them equal as we have done here for simplicity; this can therefore be considered a conservative evaluation of \lonestar.
The starred curve represents tuning $\sigmatx^2 = \sigmarx^2$ at each $\inrrxbar$ to maximize sum spectral efficiency.
At very low $\inrrxbar$, full-duplex operation ``is free'', as transmit and receive beams couple sufficiently low self-interference on their own.
In this region, minimizing the coverage variance $\sigmatx^2 = \sigmarx^2$ is ideal since it preserves near-maximum $\snrtx$ and $\snrrx$.
As $\inrrxbar$ increases, \textbf{increasing $\sigmatx^2 = \sigmarx^2$ supplies the system with greater robustness to self-interference}, evidenced by the rightward shift of the curves.
When $\inrrxbar$ is overwhelmingly high, \lonestar cannot mitigate self-interference enough, succumbing to self-interference and offering at most around $\sesumnorm = 0.5$ (essentially half-duplexing).
Notice that in this region, it is optimal to return to using very low  $\sigmatx^2 = \sigmarx^2$ in order to preserve half-duplex performance at worst by maximizing \gsnr.
Clearly, choosing $\sigmatx^2 = \sigmarx^2$ is an important design decision and depends heavily on system factors (chiefly $\inrrxbar$).
In practice, we imagine a system relying on a lookup table to retrieve the optimal $\sigmatx^2$ and $\sigmarx^2$ based on its measured $\inrrxbar$ (along with other system factors), which would likely depend on simulation and experimentation. 
Henceforth, we assume that the optimal $\sigmatxrxsq$ is used which maximizes $\sesumnorm$.


\section{Comparing \lonestar Against Conventional Beamforming Codebooks} \label{sec:simulation-results}

\begin{figure*}
    \centering
    \subfloat[As a function of link quality, $\snrtxbar = \snrrxbar$.]{\includegraphics[width=\linewidth,height=0.265\textheight,keepaspectratio]{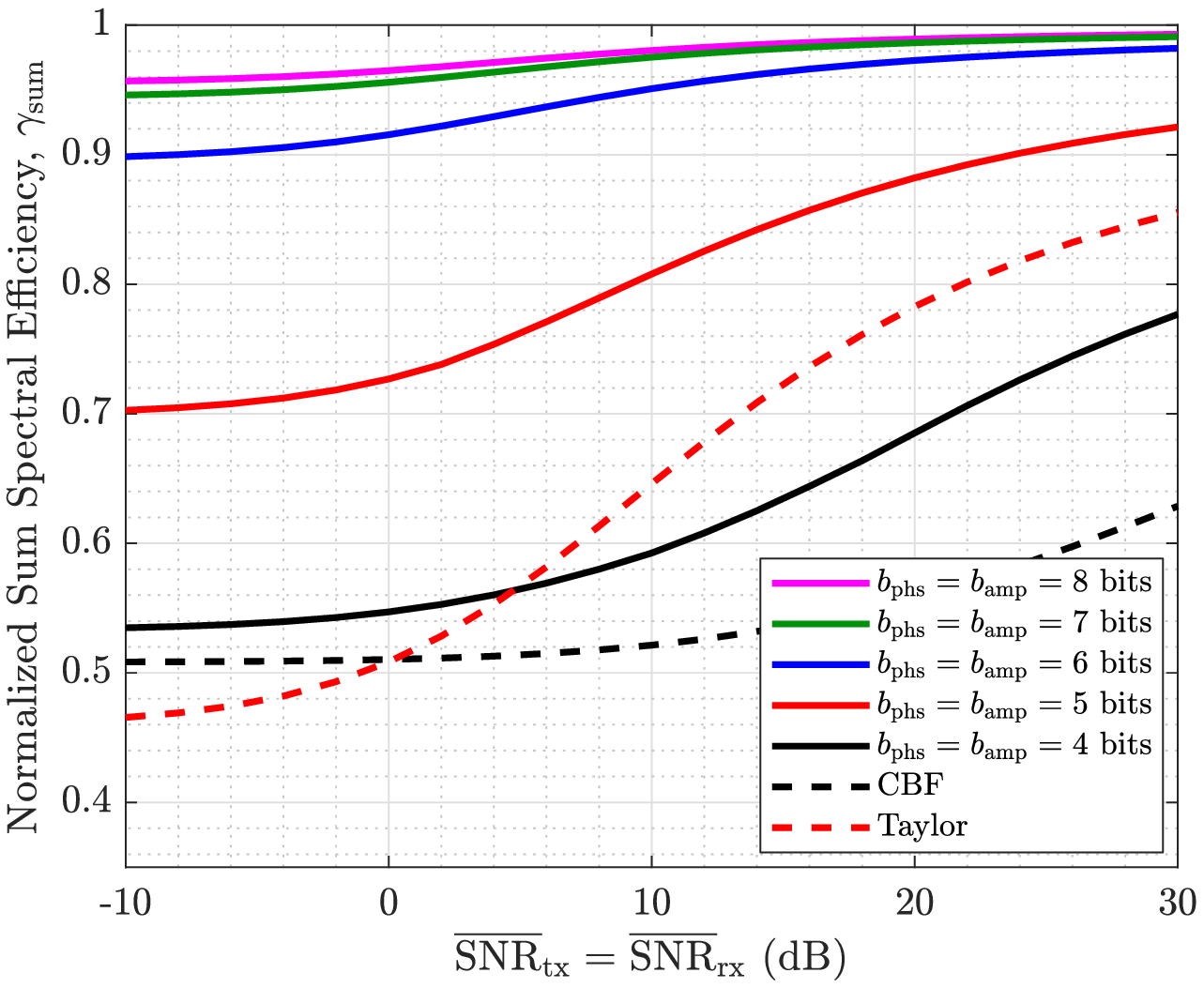}
        \label{fig:compare-a}}
    \quad
    \subfloat[As a function of self-interference strength, $\inrrxbar$.]{\includegraphics[width=\linewidth,height=0.261\textheight,keepaspectratio]{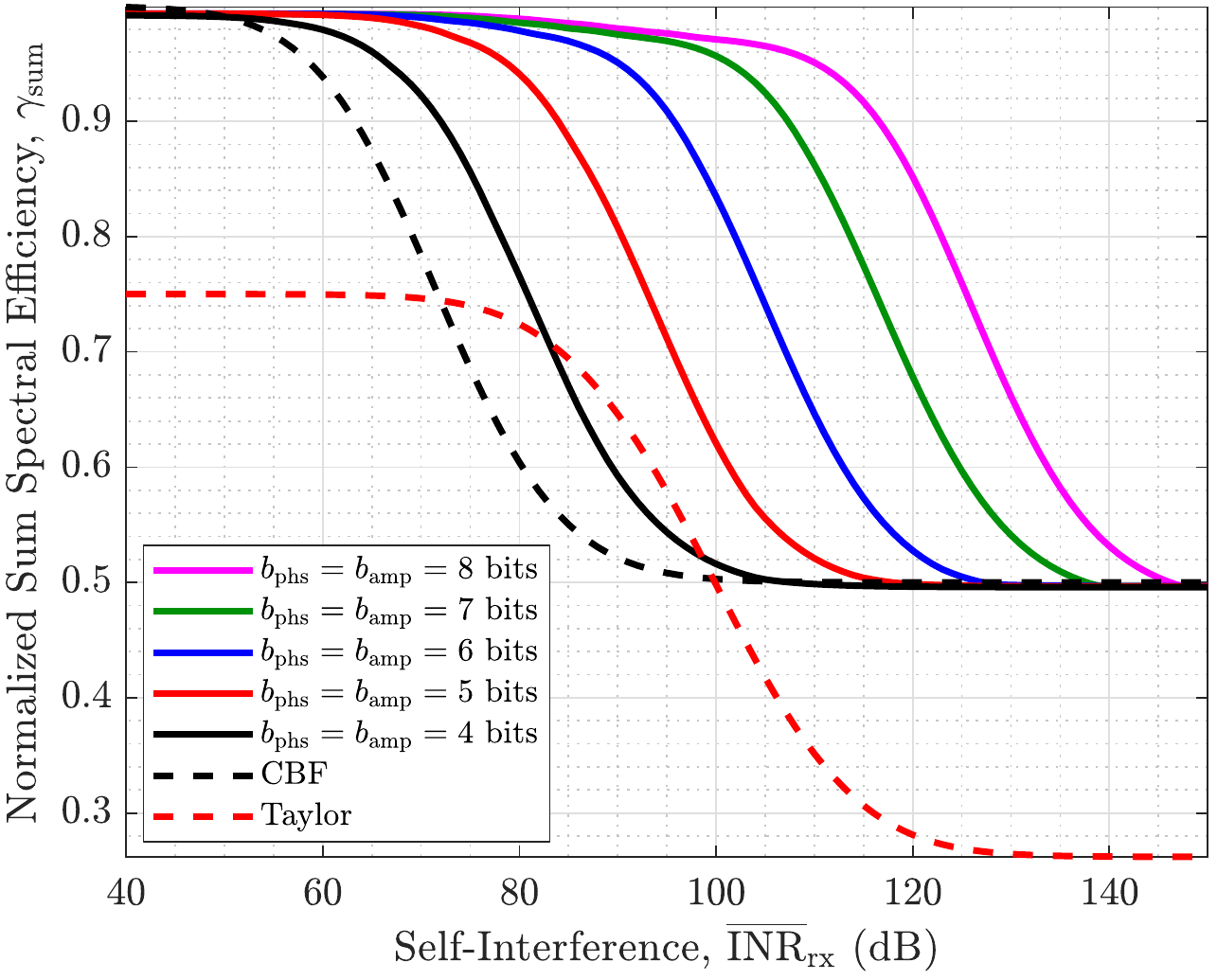}
        \label{fig:compare-b}}
    \caption{Normalized sum spectral efficiency $\sesumnorm$ (a) as a function of $\snrtxbar = \snrrxbar$ where $\inrrxbar = 90$ dB; (b) as a function of $\inrrxbar$ where $\snrtxbar = \snrrxbar = 10$ dB. In both, $\inrtx = -\infty$ dB, $\epsilon^2 = -\infty$ dB, and $\sigmatx^2 = \sigmarx^2$ are tuned to maximize sum spectral efficiency. \cbf and \taylor codebooks are each with $\bitsphaseamp = 8$ bits.}
    \label{fig:compare}
\end{figure*}

Having presented the simulation setup and performance metrics in the previous section, we now compare performance with \lonestar versus that with conventional codebooks.
In \figref{fig:compare}, we compare \lonestar codebooks against our two baseline codebooks: the \cbf and \taylor codebooks. 
We assume no cross-link interference ($\inrtx = -\infty$ dB) and no channel estimation error ($\eps^2 = -\infty$ dB); we will evaluate \lonestar against both of these shortly.
In \figref{fig:compare-a}, we plot the normalized sum spectral efficiency $\sesumnorm$ achieved by \lonestar as a function of $\snrtxbar = \snrrxbar$ for various resolutions $\bitsphase = \bitsamp$, where $\inrrxbar = 90$ dB\footnote{We commonly use $\inrrxbar = 90$ dB since this produces a median $\inrrx$ with the \cbf codebook that aligns with our \ginr measurements using an actual $28$ GHz phased array platform in \cite{roberts_att_angular}.}. 
The dashed black curve denotes the \cbf codebook and the dashed red curve denotes the \taylor codebook, both of which are configured with $\bitsphaseamp = 8$ bits.
At low $\snrtxrxbar$, the \cbf codebook can deliver $\sesumnorm \approx 0.5$, since the receive link is overwhelmed by self-interference while the transmit link sees nearly full beamforming gain.
The \taylor codebook sacrifices beamforming gain to reduce its side lobes and therefore achieves only $\sesumnorm \approx 0.46$ at low $\snrtxrxbar$, suggesting that it is also plagued by self-interference at this particular $\inrrxbar$, even with reduced side lobes.
As $\snrtxrxbar$ increases, \cbf and \taylor codebooks naturally both improve their performance as the impacts of self-interference diminish with increased \gsnr.
Shown as solid lines, \lonestar improves significantly as $\bitsphaseamp$ increases from $4$ bits to $8$ bits. 
Even at low $\snrtxrxbar$, \textbf{\lonestar can deliver sum spectral efficiencies well above that of conventional codebooks by better rejecting self-interference while maintaining high beamforming gain.}
This is especially true when $\bitsphaseamp \geq 6$ bits, which allows \lonestar to deliver upwards of $90$\% of the full-duplex codebook capacity.
At high $\snrtxrxbar$, \lonestar naturally improves, saturating due to diminishing gains in $\logtwo{1+x}$.
For $\snrtxrxbar \geq 5$ dB, the \taylor codebook can significantly outperform the \cbf codebook and can marginally outperform \lonestar having $\bitsphaseamp = 4$ bits, especially at high \gsnr.

In \figref{fig:compare-b}, we now fix $\snrtxrxbar = 10$ dB and vary $\inrrxbar$, keeping all other factors the same as in \figref{fig:compare-a}.
\cbf and \lonestar both can nearly reach the full-duplex codebook capacity at low $\inrrxbar$, where self-interference is negligible and near maximal beamforming gain can be delivered.
The \taylor codebook sacrifices beamforming gain for reduced side lobes and therefore falls well short of $\sesumnorm = 1$ even in the presence of extremely weak self-interference. 
\textbf{As $\inrrxbar$ is increased, \cbf quickly succumbs to self-interference, whereas \lonestar proves to be notably more robust to such.}
In fact, with each unit increase in resolution $\bitsphaseamp$, \lonestar supplies around $10$ dB of added robustness to self-interference.
The \taylor codebook can outperform the \cbf codebook but only for $\inrrxbar \in \brackets{72,100}$ dB before also being overwhelmed by self-interference.
At higher \gsnr, when $\snrtxrxbar \geq 10$ dB, this $\inrrxbar$ range would widen, as evidenced by \figref{fig:compare-a}.
As $\inrrxbar$ increases to extremely high levels, \lonestar converges to offer nearly the same sum spectral efficiency as \cbf of $\sesumnorm = 0.5$, essentially only serving the downlink user due to uplink being overwhelmed by self-interference.
Each full-duplex \mmwave \bs will see a unique $\inrrxbar$ depending on a variety of parameters including transmit power, noise power, and isolation between its transmit and receive arrays.
This makes \lonestar an attractive codebook design since it is more robust than conventional codebooks over a wide range of $\inrrxbar$ and provides greater flexibility when designing full-duplex \mmwave platforms.
With \lonestar, systems can operate at higher $\inrrxbar$, meaning they can potentially increase transmit power without being severely penalized by increased self-interference, for instance.


\subsection{For Various Link Qualities, $\parens{\snrtxbar,\snrrxbar}$}

\begin{figure*}
    \centering
    \subfloat[Taylor codebook.]{\includegraphics[width=0.475\linewidth,height=\textheight,keepaspectratio]{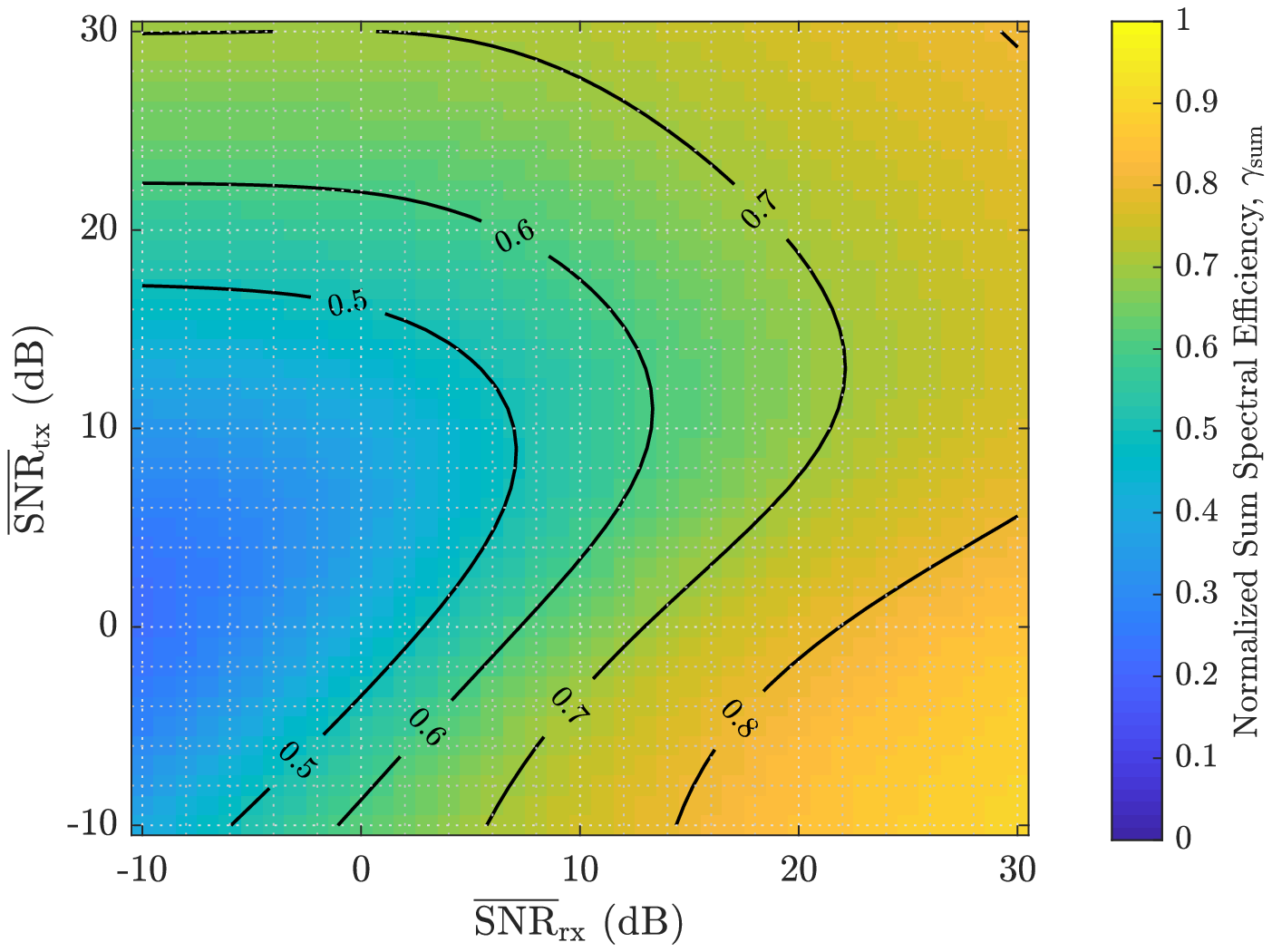}
        \label{fig:heatmap-snr-a}}
    \quad
    \subfloat[\lonestar codebooks.]{\includegraphics[width=0.475\linewidth,height=\textheight,keepaspectratio]{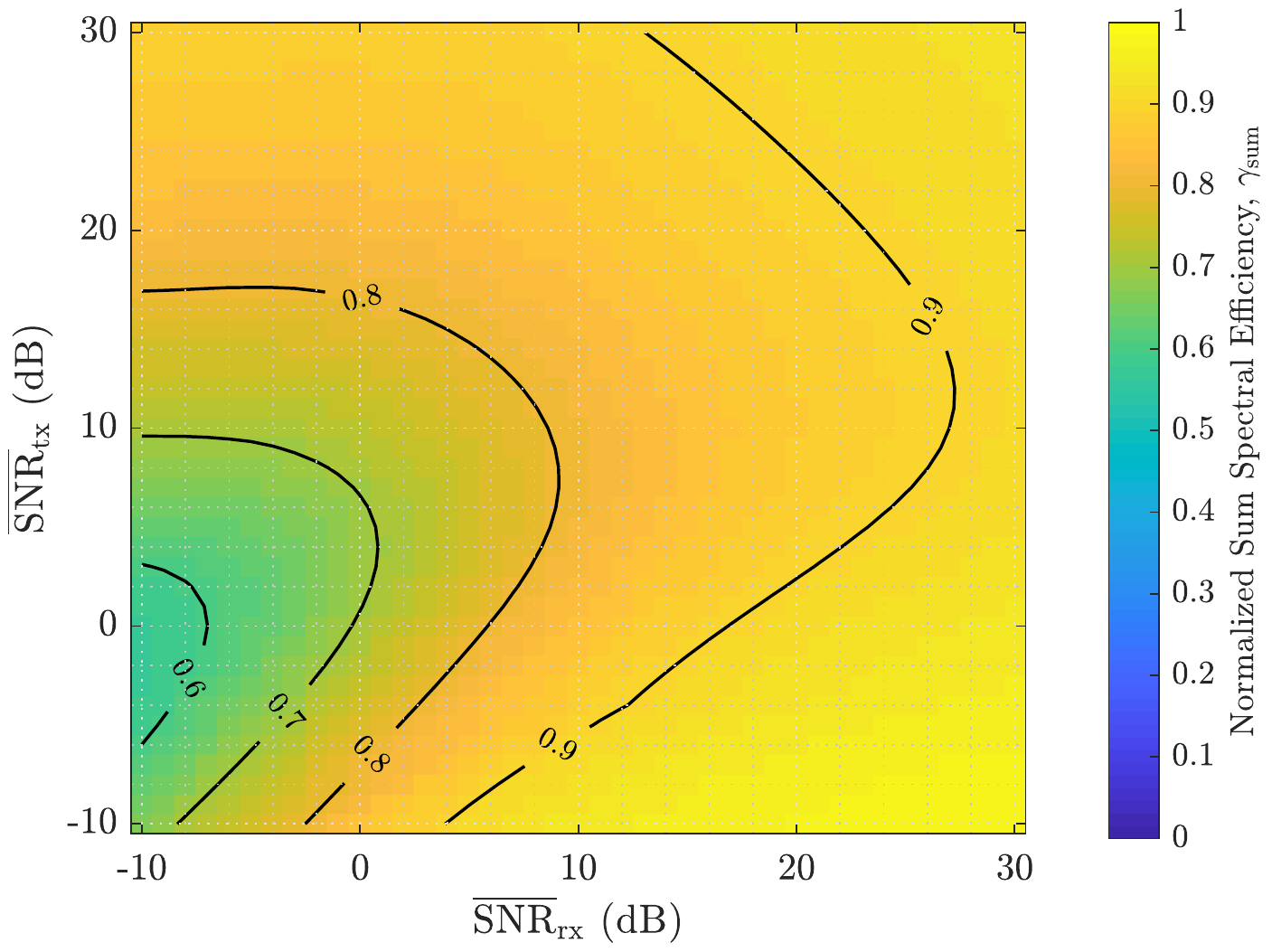}
        \label{fig:heatmap-snr-b}}
    \caption{Normalized sum spectral efficiency $\sesumnorm$ as a function of $\parens{\snrtxbar,\snrrxbar}$ for the (a) Taylor codebook ($\bitsphaseamp = 8$ bits) and (b) \lonestar codebooks, where $\inrrxbar = 90$ dB, $\inrtx = -\infty$ dB, $\bitsphaseamp = 6$ bits, and $\eps^2 = -\infty$ dB. \lonestar delivers higher $\sesumnorm$ broadly across \gpsnr and demands lower \gpsnr to net $\sesumnorm \geq 0.5$.}
    \label{fig:heatmap-snr}
\end{figure*}

Rather than assume symmetric downlink/uplink quality $\snrtxrxbar$, we now consider $\snrtxbar \neq \snrrxbar$.
In \figref{fig:heatmap-snr-a}, we show the normalized sum spectral efficiency achieved by the \taylor codebook as a function of $\parens{\snrtxbar,\snrrxbar}$, where $\inrrxbar = 90$ dB and $\bitsphaseamp = 8$ bits.
In \figref{fig:heatmap-snr-b}, we show that of \lonestar with $\bitsphaseamp = 6$ bits. 
We chose to compare against the \taylor codebook rather than the \cbf codebook because it better competes with \lonestar at $\inrrxbar = 90$ dB (see \figref{fig:compare-a}).
Clearly, \textbf{\lonestar can deliver higher spectral efficiencies} than the \taylor codebook broadly over $\parens{\snrtxbar,\snrrxbar}$.
It is important to notice that the region where $\sesumnorm \geq 0.5$---where full-duplexing is superior to half-duplexing with equal \tdd---grows significantly with \lonestar versus the \taylor codebook.
This means that full-duplex operation is justified over a much broader range of $\parens{\snrtxbar,\snrrxbar}$ with \lonestar compared to with conventional codebooks.
It is also important to notice that \textbf{\lonestar introduces a significant \gsnr gain versus conventional codebooks}, since it demands a much lower $\parens{\snrtxbar,\snrrxbar}$ to achieve a desired sum spectral efficiency $\sesumnorm$.
Note that we have tuned $\sigmatxrxsq$, meaning the performance of \lonestar may improve if $\sigmatxsq$ and $\sigmarxsq$ were tuned separately, especially with highly asymmetric \gpsnr.

\subsection{For Various Levels of Interference, $\parens{\inrtx,\inrrxbar}$}

\begin{figure*}
    \centering
    \subfloat[Taylor codebook.]{\includegraphics[width=0.475\linewidth,height=\textheight,keepaspectratio]{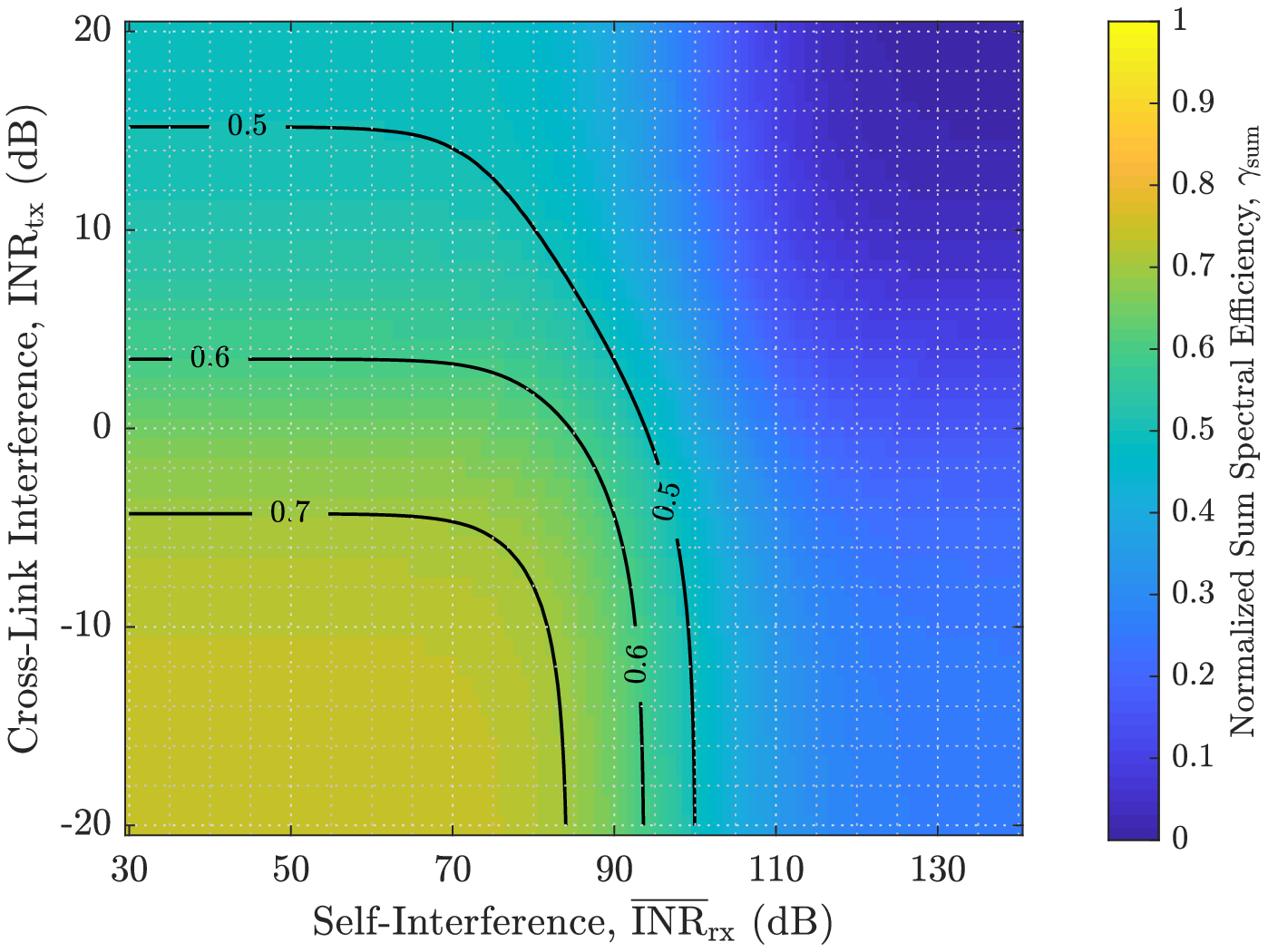}
        \label{fig:heatmap-inr-a}}
    \quad
    \subfloat[\lonestar codebooks.]{\includegraphics[width=0.475\linewidth,height=0.255\textheight,keepaspectratio]{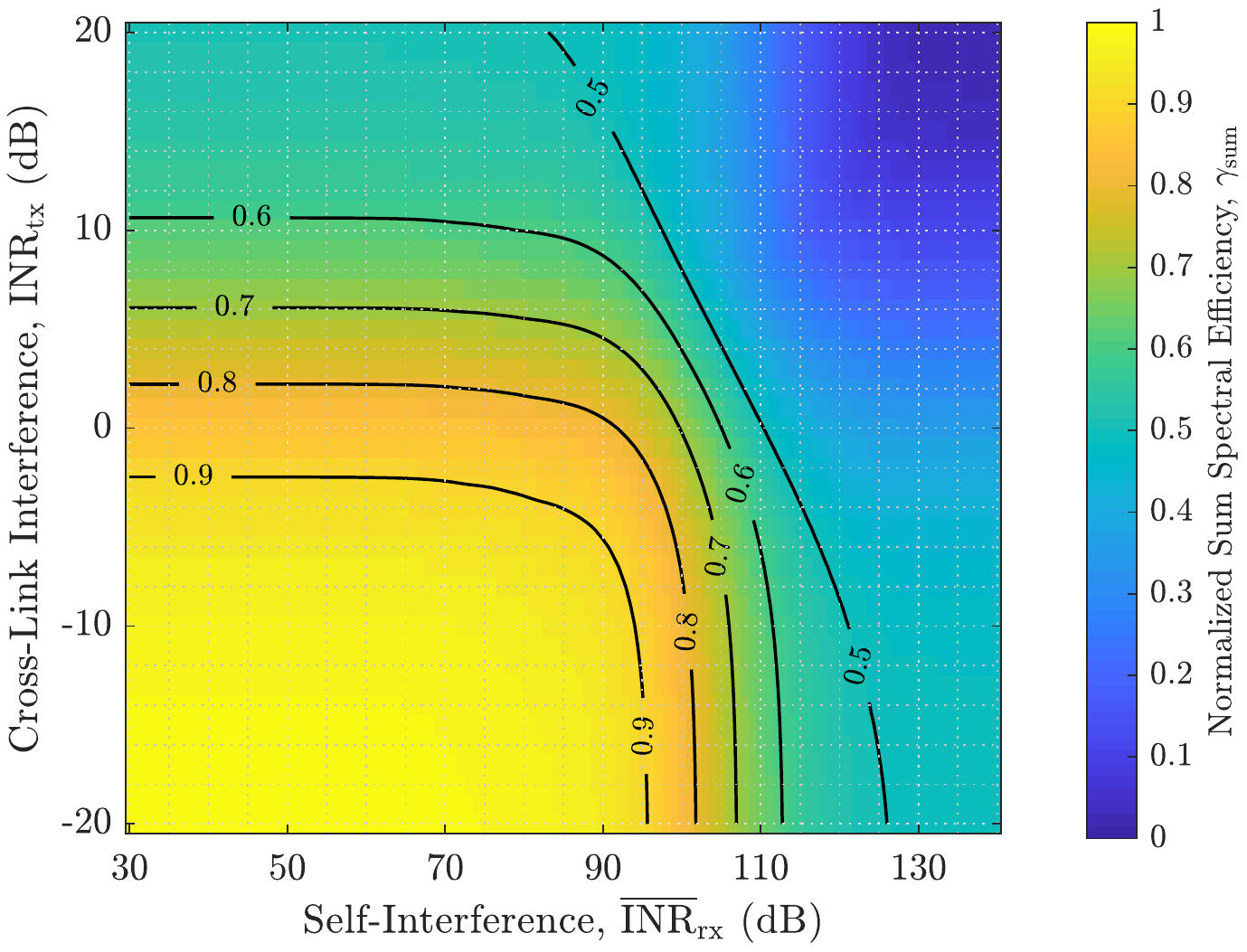}
        \label{fig:heatmap-inr-b}}
    \caption{Normalized sum spectral efficiency $\sesumnorm$ as a function of $\parens{\inrtx,\inrrxbar}$ for the (a) Taylor codebook, where $\bitsphaseamp = 8$ bits; and (b) \lonestar codebooks, where $\bitsphaseamp = 6$ bits. In both, $\snrtxrxbar = 10$ dB and $\eps^2 = -\infty$ dB. \lonestar is more robust to self-interference and cross-link interference than conventional codebooks.}
    \label{fig:heatmap-inr}
\end{figure*}

Similar to \figref{fig:heatmap-snr}, we now compare the \taylor codebook against \lonestar for various $\parens{\inrtx,\inrrxbar}$ in \figref{fig:heatmap-inr}, where we have fixed $\snrtxrxbar = 10$ dB.
Again, \lonestar can clearly deliver much higher spectral efficiencies than the \taylor codebook, achieving nearly the full-duplex codebook capacity when cross-link interference and self-interference are both low.
Even at very low \gpinr, the \taylor codebook cannot approach $\sesumnorm = 1$ since it sacrificed beamforming gain for side lobe suppression.
At high \gpinr, both \lonestar and the \taylor codebook struggle to deliver high $\sesumnorm$, largely due to the fact that neither can reduce cross-link interference.
In between, at moderate \gpinr, \lonestar can sustain spectral efficiencies notably higher than the \taylor codebook.
\textbf{\lonestar's added robustness to cross-link interference and self-interference is on display here}, as it can tolerate higher $\parens{\inrtx,\inrrxbar}$ than the \taylor codebook while delivering the same $\sesumnorm$.
This is thanks to \lonestar's ability to net a higher receive link spectral efficiency, which it may sacrifice in the presence of high cross-link interference $\inrtx$.
At an $\inrrxbar = 50$ dB, for example, \lonestar provides around $10$ dB of robustness to $\inrtx$ for $\sesumnorm = 0.7$, compared to the \taylor codebook.
At an $\inrtx = -10$ dB, \lonestar provides around $30$ dB of robustness to self-interference $\inrrxbar$ for $\sesumnorm = 0.7$.

\subsection{Impact of Channel Estimation Error and Channel Structure}

\begin{figure*}
    \centering
    \subfloat[Channel estimation error.]{\includegraphics[width=0.475\linewidth,height=0.255\textheight,keepaspectratio]{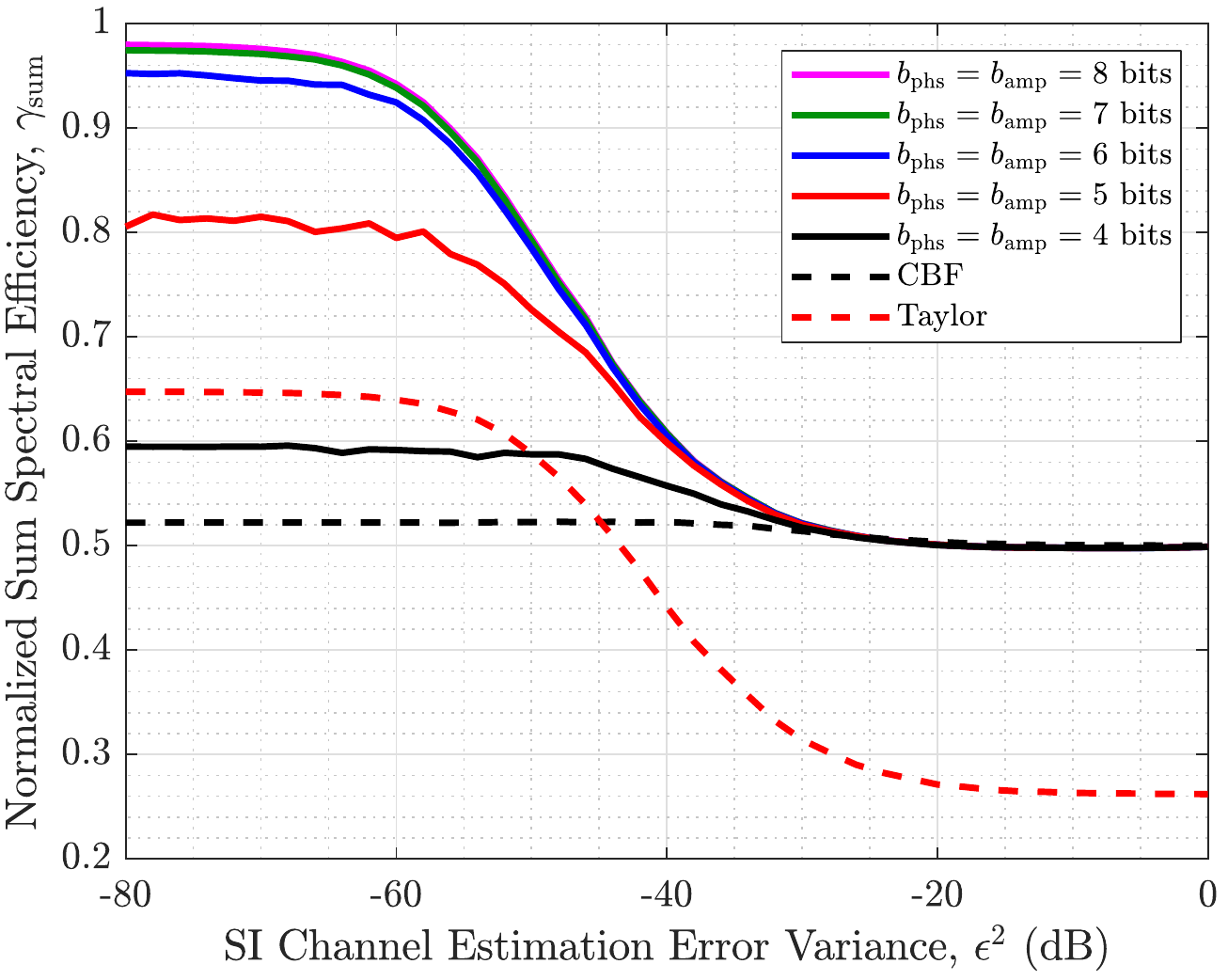}
        \label{fig:mixture-error-a}}
    \quad
    \subfloat[Gaussian mixing.]{\includegraphics[width=0.475\linewidth,height=0.255\textheight,keepaspectratio]{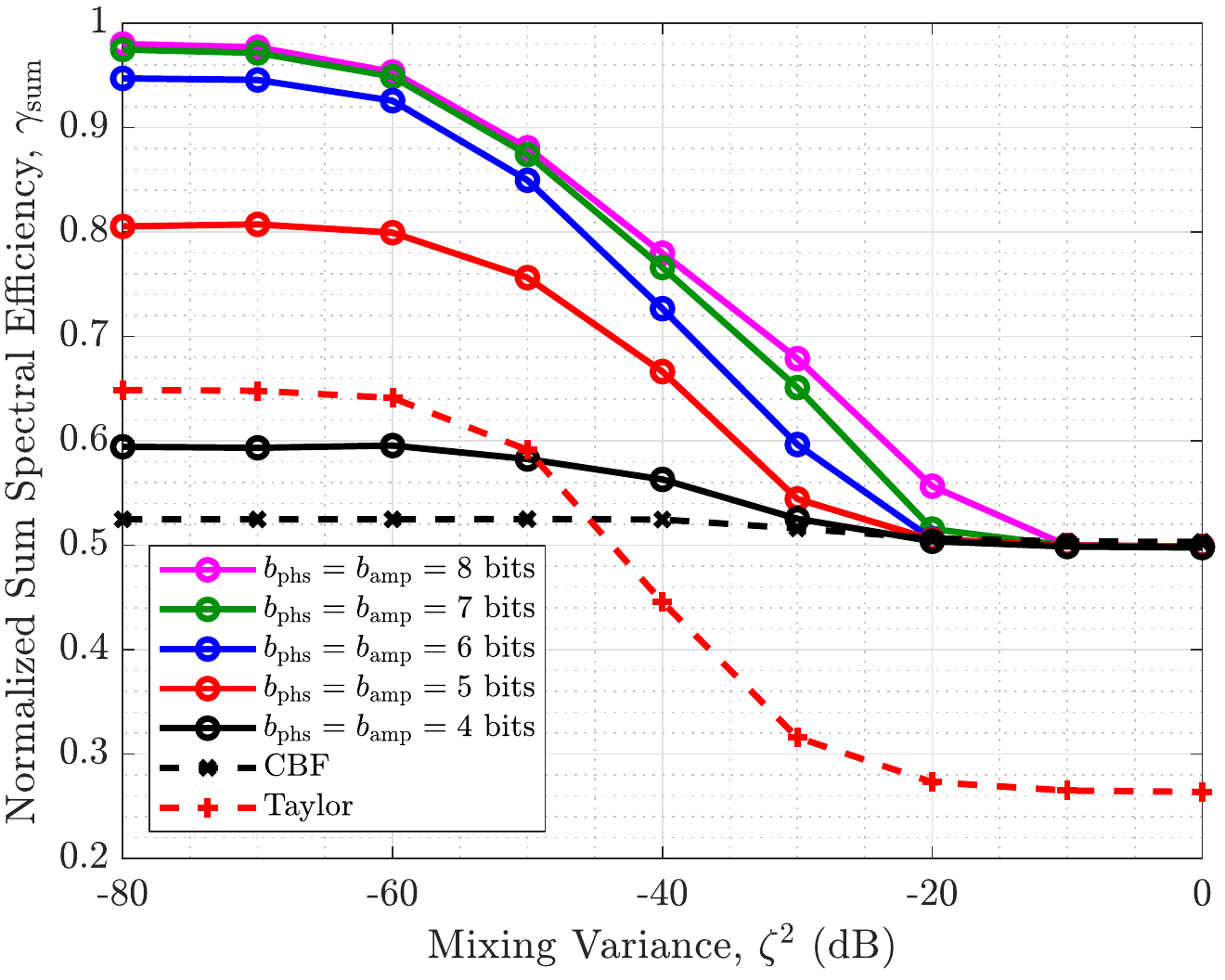}
        \label{fig:mixture-error-b}}
    \caption{Normalized sum spectral efficiency as a function of (a) self-interference channel estimation error variance $\eps^2$ and (b) mixing variance $\zeta^2$. In both, $\snrtxbar = \snrrxbar = 10$ dB, $\inrrxbar = 90$ dB, and $\inrtx = -\infty$ dB. The true channel $\mH$ is distributed the same in both, but in (b), we assume perfect channel knowledge, $\eps^2 = -\infty$ dB.}
    \label{fig:mixture-error}
\end{figure*}

Now, in \figref{fig:mixture-error-a}, we examine the impacts of imperfect self-interference channel knowledge, $\eps^2 > -\infty$ dB.
To do so, we model $\mHbar$ as in \eqref{eq:spherical-wave}, meaning the true channel $\mH$ is Gaussian distributed about $\mHbar$ according to \eqref{eq:channel}.
We chose to model it in this fashion, rather than use $\mH$ as the spherical-wave model in \eqref{eq:spherical-wave}, for two reasons.
First, it is more challenging for \lonestar to face a channel under this mixed model (as we will see shortly in \figref{fig:mixture-error-b}).
Second, it represents the potential case where the estimate $\mHbar$ is computed directly based on the relative geometry of the transmit and receive arrays using the spherical-wave model (which can be done \textit{a priori}) and the estimation error captures deviations from this assumed spherical-wave behavior.
In \figref{fig:mixture-error-a}, we plot the normalized sum spectral efficiency as a function of self-interference channel estimation error variance $\epsilon^2$.
With accurate estimation of $\mH$, \lonestar performs quite well as it has proven thus far.
As the quality of estimation degrades, \lonestar has difficulty reliably avoiding self-interference.
Its performance suffers but remains well above $\sesumnorm = 0.5$ before converging to performance offered by the \cbf codebook around $\eps^2 = -30$ dB.
The \cbf codebook sees mild degradation with increased estimation error.
Clearly, the \taylor codebook, like \lonestar, suffers as $\eps^2$ increases. 
It may seem strange that the \cbf and \taylor codebooks both vary as a function of estimation error $\eps^2$ since neither codebook depends on self-interference channel knowledge. 
This can be explained by the fact that the estimation error $\mDelta$ is distributed differently than the estimate $\mHbar$.
In other words, the \taylor codebook is inherently more robust to the more structured spherical-wave self-interference channel than a Gaussian one.
\textbf{Gaussian-distributed error proves to be especially difficult for \lonestar and \taylor to endure.} 
As such, it is evident that imperfect channel knowledge \textit{and} the fact that estimation error is Gaussian both contribute to \lonestar's degradation as $\eps^2$ increases.

We dissect the impacts of estimation error versus that of mixing the spherical-wave model with a Gaussian matrix in \figref{fig:mixture-error-b} by considering the mixed self-interference channel model
\begin{align}
\mH = \frac{\mH_{\mathrm{SW}} + \mH_{\mathrm{Ray}}}{\normfro{\mH_{\mathrm{SW}} + \mH_{\mathrm{Ray}}}} \cdot \sqrt{\Nt\cdot\Nr}, \qquad \entry{\mH_{\mathrm{Ray}}}{m,n} \sim \distcgauss{0}{\zeta^2} \ \forall \ m,n \label{eq:mixture}
\end{align}
where $\mH_{\mathrm{SW}}$ is the spherical-wave model defined in \eqref{eq:spherical-wave} and $\mH_{\mathrm{Ray}}$ is Rayleigh faded with variance $\zeta^2$.
The scaling in \eqref{eq:mixture} simply ensures $\normfro{\mH}^2 = \Nt\cdot\Nr$.
We evaluate \lonestar using this channel model for various $\zeta^2$, which we term the \textit{mixing variance}.
In doing so, we assume perfect knowledge of the realized channel $\mH$ (i.e., $\eps^2 = -\infty$ dB) to solely investigate the effects of this mixing.
The spherical-wave channel $\mH_{\mathrm{SW}}$ is a highly structured channel based on the relative geometry between the transmit and receive arrays at the \bs.
The Rayleigh faded channel $\mH_{\mathrm{Ray}}$ (a Gaussian matrix) lay at the other extreme, lacking any inherent structure.
It is important to notice that the distribution of true channel $\mH$ here is identical to that examined in \figref{fig:mixture-error-a}.
In this case, however, we consider $\eps^2 = -\infty$ dB to investigate the impacts of $\mH$ becoming increasingly Gaussian, rather than those stemming from imperfect channel knowledge.

In \figref{fig:mixture-error-b}, we plot the normalized sum spectral efficiency as a function of the mixing variance $\zeta^2$. 
Having considered perfect channel knowledge $\eps^2 = -\infty$ dB here, \figref{fig:mixture-error-b} is an upper bound on \figref{fig:mixture-error-a} since they follow the same channel model.
At very low $\zeta^2$, $\mH \approx \mH_{\mathrm{SW}}$, allowing \lonestar to achieve appreciable spectral efficiencies above conventional codebooks, as we have seen thus far. 
As $\zeta^2$ increases, the sum spectral efficiency offered by \lonestar tends to decrease, though so does the \taylor codebook and the \cbf codebook slightly so.
At high $\zeta^2$, \lonestar suffers since it cannot find ways to pack the transmit and receive beams such that they do not couple across the self-interference channel.
This is intimately related to the fact that Gaussian matrices (almost surely) tend to inflict self-interference across all dimensions of $\mH$, making it a statistically more difficult channel for \lonestar to build codebooks that minimize transmit-receive beam coupling.
\textbf{Thus, even with perfect channel estimation, \lonestar struggles with self-interference channels that are heavily Gaussian.}
From the results in \figref{fig:mixture-error}, we observe that there is the need for accurate and perhaps frequent self-interference channel estimation.
Our hope is that this can be accomplished by taking advantage of the strength of the self-interference channel, along with a potentially known channel structure and occasional---but thorough---measurements at the \bs, as mentioned in \secref{sec:problem-formulation}.

\section{Conclusion and Future Directions} \label{sec:conclusion}

\edit{
Conventional beamforming codebooks are not suitable for full-duplex \mmwave systems since they do not offer much robustness to self-interference.
Meanwhile, existing beamforming-based solutions for full-duplex \mmwave systems typically do not support codebook-based beam alignment, demand real-time knowledge of high-dimensional downlink/uplink \mimo channels, and do not scale well when serving many users over time---all of which are practical shortcomings.
We have presented \lonestar, a novel approach to enable full-duplex \mmwave communication systems through the design of analog beamforming codebooks that are robust to self-interference.
Our design does not rely on downlink/uplink channel knowledge and therefore need only be executed once per coherence time of the self-interference channel, allowing the same \lonestar codebooks to be used for beam alignment and to serve many downlink-uplink user pairs thereafter.
We illustrate that \lonestar can mitigate self-interference to levels sufficiently low for appreciable spectral efficiency gains compared to conventional codebooks by being more robust to self-interference and cross-link interference broadly across downlink and uplink \gpsnr.}

\edit{
This work has motivated a variety of future work for full-duplex \mmwave systems.
Extensive study and modeling of \mmwave self-interference channels and their coherence time will be important to understanding the potentials of \lonestar---and full-duplex \mmwave systems in general---along with developing efficient means for self-interference channel estimation.
Also, methods to dynamically update \lonestar as the self-interference channel drifts would be valuable future work, along with implementations of \lonestar using real \mmwave platforms.
Naturally, there are other ways to improve upon this work, such as by better handling limited phase and amplitude control and by creating efficient, dedicated solvers for \lonestar.
Beyond \lonestar, measurement-driven beamforming-based full-duplex solutions (potentially using machine learning) that do not require self-interference \mimo channel knowledge (e.g., \steer \cite{roberts_steer}) would be excellent contributions to the research community.
}





\section*{Acknowledgments}

We would like to thank Marius Arvinte and Ruichen Jiang from the University of Texas at Austin for their discussions related to solving our design optimization problem.
I.~P.~Roberts is supported by the National Science Foundation Graduate Research Fellowship Program under Grant No.~DGE-1610403. 
Any opinions, findings, and conclusions or recommendations expressed in this material are those of the authors and do not necessarily reflect the views of the National Science Foundation.



\bibliographystyle{bibtex/IEEEtran}
\bibliography{bibtex/IEEEabrv,refs}

\end{document}